\pdfoutput=1
\documentclass[aps,prd,10pt,preprintnumbers,showpacs,twocolumn,
superscriptaddress , floatfix,nofootinbib]{revtex4-1}

\usepackage{latexsym} 
\usepackage{amssymb} 
\usepackage{amsmath} 
\usepackage{amsfonts}
\usepackage{bm}
\usepackage{color}
\usepackage{times} 
\usepackage{units}
\usepackage{hyperref}
\usepackage{tabularx}
\usepackage[utf8x]{inputenc}
\usepackage{graphicx} 
\usepackage[squaren]{SIunits} 
\usepackage[inline]{enumitem}
\usepackage{xspace}

\graphicspath{ {../plots/} }

\newcommand{\Fref}[1]{Fig.~\ref{#1}} 
\newcommand{\Sref}[1]{Sec.~\ref{#1}} 
\newcommand{\Tref}[1]{Table~\ref{#1}}

\begin{document}

\title{Numerical Inside View of Hypermassive Remnant Models for GW170817}

\author{W. Kastaun}
\author{F. Ohme}

\affiliation{Max Planck Institute for Gravitational Physics
(Albert Einstein Institute), Callinstr. 38, D-30167 Hannover, Germany}
\affiliation{Leibniz Universität Hannover, D-30167 Hannover, Germany}

\def\ECCMNONEQ{\ensuremath{0.009}\xspace}
\def\ECCMEQUAL{\ensuremath{0.010}\xspace}
\def\SNRMERGQ09O2{\ensuremath{1.1}\xspace}
\def\SNRMERGQ09aLIGO{\ensuremath{4.8}\xspace}
\def\SNRMERGQ09Voyager{\ensuremath{1.0}\xspace}
\def\SNRMERGQ09ET{\ensuremath{2.6}\xspace}
\def\SNRMERGQ10O2{\ensuremath{1.1}\xspace}
\def\SNRMERGQ10aLIGO{\ensuremath{4.9}\xspace}
\def\SNRMERGQ10Voyager{\ensuremath{1.0}\xspace}
\def\SNRMERGQ10ET{\ensuremath{2.6}\xspace}
 \def\MFBBHMEQUAL{\ensuremath{2.60}\xspace}
\def\MFBBHMNONEQ{\ensuremath{2.60}\xspace}
\def\CHIFBBHMEQUAL{\ensuremath{0.69}\xspace}
\def\CHIFBBHMNONEQ{\ensuremath{0.68}\xspace}

\begin{abstract}
The first multimessenger observation attributed to a merging neutron star binary
provided an enormous amount of observational data. Unlocking the full potential
of this data requires a better understanding of the merger process and the early post-merger
phase, which are crucial for the later evolution that eventually leads to observable 
counterparts. In this work, we perform standard hydrodynamical numerical simulations 
of a system compatible with GW170817. We focus on a single equation of state (EOS) and
two mass ratios, while neglecting magnetic fields and neutrino radiation. We then apply 
newly developed postprocessing and visualization techniques to the results obtained for 
this basic setting. The focus lies on understanding the three-dimensional structure of 
the remnant, most notably the fluid flow pattern, and its evolution until collapse.
We investigate the evolution of mass and angular momentum distribution up to collapse,
as well as the differential rotation along and perpendicular to the equatorial plane.
For the cases that we studied, the remnant cannot be adequately modeled as a 
differentially rotating axisymetric NS. Further, the dominant aspect leading to collapse 
is the GW radiation and not internal redistribution of angular momentum. We relate  
features of the gravitational wave signal to the evolution of the merger remnant, and 
make the waveforms publicly available. Finally, we find that the three-dimensional 
vorticity field inside the disk is dominated by medium-scale perturbances and not the  
orbital velocity, with potential consequences for magnetic field amplification effects. 
\end{abstract}


\maketitle

\section{Introduction}
\label{sec:intro}

This work is motivated by the first multi-messenger detection compatible
with the coalescence of two neutron stars (NSs). 
The gravitational wave (GW) event 
GW170817 detected by the LIGO/Virgo observatories matches the inspiral 
of two compact objects in the NS mass range 
\cite{LVC:BNSDetection,LVC:BNSSourceProp:2019}.  
After a delay around $1.7\usk\second$, the GW signal was followed by 
short gamma ray burst (SGRB) event GRB170817A observed by Fermi and INTEGRAL 
satellites and attributed to the same source \cite{LVC:GWGRB:2017}. 
Later observations also revealed radio signals \cite{Mooley2018b,Ghirlanda2019} 
that likely correspond to the radio-afterglow of the SGRB. 
The coincident GW and SGRB events triggered a large observational followup 
campaign \cite{LVC:MMA:2017}. 
Observations ranging from infrared to ultraviolet revealed an optical counterpart
AT2017gfo with luminosity and spectral evolution compatible with a kilonova
\cite{LVC:MMA:2017,Kasen:2017:551, Villar:2017wcc, Cowperthwaite:2017dyu}.

The comparison of those observations to theoretical expectations requires the modeling
of many different aspects of fundamental physics, such as general relativity, 
hydrodynamics, nuclear physics, neutrino physics, and magnetohydrodynamics. Modeling all potentially 
observable electromagnetic counterparts also involves a large range of timescales ranging from 
milliseconds to years. There is however little doubt that the early evolution phase up to tens 
of milliseconds after merger is of crucial importance. This phase can be studied via 
brute force three-dimensional numerical simulations and will be the topic of this work.

For predicting the expected kilonova signal, the important input from such studies are amount, 
composition, and velocity of matter dynamically ejected to infinity and of matter ejected from 
the disk.
The latter is likely relevant since the kilonova spectral evolution is best fitted by two or 
more distinct ejecta components \cite{Kasen:2017:551, Villar:2017wcc, Cowperthwaite:2017dyu}
with masses that would be at tension with purely dynamical ejection mechanism.
Although the fraction of disk mass expelled via winds is uncertain, the total mass of the disk 
poses an upper limit.  Numerical simulations suggest that the initial disk mass depends strongly 
on the total mass of the system in comparison to the maximum NS mass for the given EOS, and on 
the mass ratio.

On timescales of 0.1 s, the evolution of the disk is strongly influenced by the interaction 
with the remnant. In case of a supra- or hypermassive NS, matter can be transported into the 
disk by different mechanisms. One is a purely hydrodynamic consequence of a complicated internal 
fluid flow inside the remnant \cite{Kastaun:2015:064027,Kastaun:2016}. Another potential mechanism 
is the amplification of magnetic fields inside the remnant and disk and the resulting pressure 
\cite{Ciolfi:2019:023005}. Until collapse, a remnant NS also irradiates disk and ejecta with neutrinos
and therefore has an impact on the composition. In particular the fraction of Lanthanides in the 
ejecta has a strong impact on the optical opacity. For those reasons, the remnant lifetime is 
important with regard to the kilonova signal.

Also with regard to the SGRB signal, the mass of the disk and the delay before black-hole (BH) formation are 
likely to be very relevant parameters. Current models for the SGRB engine require either a BH 
\cite{Shibata:2006:031102} or a magnetar  \cite{Metzger2008} embedded in a massive disk. 
The question which scenario is viable, or if both are viable, is an active field of research. 
Should it be the case that BH formation is required before the SGRB, one obtains an upper limit
on the collapse delay after merger. Given the total mass of the coalescing NSs as 
inferred from the GW signal, one obtains an upper limit on the mass which the central
NS can be sustain longer than the SGRB delay. By comparing with the maximum mass of a nonrotating
NS, a robust but not very strict constraint on the EOS was obtained from GW170817 \cite{LVC:GWGRB:2017}. 
Adding further assumptions, e.g., that the remnant NS is hypermassive, results in stricter limits 
\cite{Margalit:2017dij,Shibata:2019ctb,Ruiz:2017due,LVC:EOSModelSel:2020}. It is therefore very 
important to understand the stability criteria of the remnant.

For the simpler case of isolated uniformly rotating NS, the stability conditions are well 
understood. There is a maximum mass that depends only on the EOS. In the supramassive range, 
i.e. between the maximum masses of nonrotating and uniformly rotating NS, a minimum
angular momentum is required. On timescales ${\lesssim}0.1 \second$ however, one has to take 
into account that the remnant is not uniformly rotating and that a significant fraction of 
total mass and angular momentum can be located in the disk outside the NS remnant. 

For the important case of hypermassive
NS remnants, differential rotation is needed to prevent collapse. It is a popular assumption that
collapse is caused by the dissipation of the differential rotation. Should this be the only important 
aspect, then the collapse delay depends on the effective viscosity, which is not well constrained
as it may depend on small scale magnetic field amplification. However, numerical simulations prove 
that merger remnants emit strong GWs. Since short-lived remnants are close to collapse 
already, collapse could be triggered by a relatively small angular momentum loss. It might well be
the case that the collapse delay is mainly determined by such losses instead of viscosity, or
that both aspects are relevant.

The overall rotation profile of merger remnants, which is a key aspect for stability, 
has been studied in many numerical simulations
\cite{Ciolfi:2017:063016,Kastaun:2015:064027,Kastaun:2016,Kastaun:2017,Ciolfi:2019:023005,Endrizzi:2018,
Endrizzi:2016:164001,Hanauske:2016}. All these studies find a relatively slow rotation
of the core, and a maximum rotation rate in the outer parts of the remnant.
The typical mass distribution in the remnant core seems in fact to be similar
to that of a nonrotating NS \cite{Kastaun:2015:064027,Kastaun:2016,Ciolfi:2017:063016,Endrizzi:2018}. 
This led to the conjecture that the collapse occurs
once the remnant core density profile matches the one in the core of the maximum-mass nonrotating 
NS \cite{Ciolfi:2017:063016}. This conjecture was validated for a small number of examples 
\cite{Ciolfi:2017:063016,Endrizzi:2018} but remains unproven in general.

The works above also revealed that the fluid flow can be more complex than just axisymmetric 
differential rotation, featuring secondary vortices (see \cite{Kastaun:2016,Kastaun:2017,
Ciolfi:2017:063016}). 
However, those results are restricted to the equatorial plane, and little is know 
about the 3D structure. The analysis of fluid flow patterns in numerical 
simulations is complicated because early merger phase is not fully stationary. Remnants 
show strong oscillations and can undergo a drift of compactness and rotation rate within 
few dynamical timescales. A further difficulty arises from the coordinate choices in numerical
simulations, which are not well suited for studying the remnant shape \cite{Kastaun:2015:064027}.

In this work, we focus on studying the three-dimensional structure of merger remnants obtained 
when including only the most basic ingredient of general relativistic hydrodynamics, while 
neglecting magnetic fields, effective magnetic viscosity, and neutrino radiation transport. We will analyze 
the outcome of two simulations compatible with GW170817 in depth. For this, we develop
novel postprocessing and visualization methods. We also investigate the evolution of 
the angular momentum distribution, using different measures.

For GW170817, all useful information about the postmerger phase comes from the optical 
counterparts. No GW signal could be detected after merger \cite{LVC:PM:2017,LVC:PMLong:2019}.
Future observations of similar events with third-generation GW antennas might also include direct
detection of a postmerger GW signal or strict upper limits. In order to support
the development of postmerger-GW data analysis methods, we make the waveforms extracted from 
our simulations publicly available \cite{zenodo:4570825} as a qualitative example.

\section{Methods}
\label{sec:meth}
\subsection{Evolution}
\label{sec:evol}
The general relativistic hydrodynamic equations are evolved numerically using the 
code described in \cite{Galeazzi:2013:64009,Alic:2013:64049}. The code utilizes a
finite-volume high resolution shock capturing scheme in conjunction with the 
HLLE approximate Riemann solver and the piecewise parabolic method for reconstructing 
values at the cell interfaces.  
We neither include magnetic fields nor neutrino radiation, and the electron fraction 
is passively advected along with the fluid. 
Our numerical evolution employs a standard artificial atmosphere scheme, 
with zero velocity, lowest available temperature, and a spatially constant density cut
of $6\times 10^5 \usk\gram\per\centi\meter\cubed$.

The matter equation of state is computed using a three-dimensional interpolation 
table, where the independent variables are density, temperature, and electron fraction.
For the simulations in this work, we employ the SFHO EOS \cite{Hempel:2009mc}, 
which incorporates thermal and composition effects. The EOS was taken from the 
CompOSE EOS collection \cite{ComposeEOS}. The table only contains temperatures 
above $0.1\usk \mega\electronvolt$. In the context of a binary NS (BNS) merger 
simulation, this is not problematic 
since the thermal pressure at this temperature is negligible to the degeneracy pressure
except for very low densities. Although dynamically ejected matter becomes diluted, 
it is also very hot and therefore not affected. 

The spacetime is evolved using the \texttt{McLachlan} code \cite{Brown:2009:44023},
which is part of the Einstein Toolkit \cite{Loeffler:2012:115001}. This code implements two 
formulations of the evolution equation: the BSSN formulation 
\cite{Nakamura:1987:1,Shibata:1995:5428,Baumgarte:1998:24007}, 
and the newer conformal and spatially covariant Z4 evolution scheme described 
in \cite{Alic:2012:64040, Alic:2013:64049}. Here we use the latter because
of its constraint damping capabilities.
We employ standard gauge conditions, choosing the lapse according to the 
$1+\log$-slicing condition \cite{Bona:1995:600} and the shift vector according to
the hyperbolic $\Gamma$-driver condition \cite{Alcubierre:2003:84023}. 
At the outer boundary, we use the Sommerfeld radiation boundary condition.

The time integration of the coupled hydrodynamic and spacetime evolution
equations is carried out using the Method of Lines (MoL) with a 4th-order 
Runge-Kutta scheme. Further, we use Berger-Oliger moving-box mesh refinement 
provided by the \texttt{Carpet} code \cite{Schnetter:2004:1465}. 
For tests of the code, we refer the reader to \cite{Galeazzi:2013:64009,Alic:2013:64049}.

In total, we use six refinement 
levels, each of which has twice the resolution of the next finer one. 
The four coarsest levels consist of a simple hierarchy of nested cubes centered
around the origin. The two finest levels consist of nested cubes that follow each 
of the stars during inspiral. 
Near merger, those are replaced by non-moving nested cubes centered around the 
origin. The finest grid spacing is $221 \usk\meter$. The outer boundary is located 
at $950\usk\kilo\meter$, and the finest level after merger covers a radius of 
$28\usk\kilo\meter$. Finally, we use reflection symmetry across the orbital plane.

\subsection{Initial data}
\label{sec:models}

In this work, we evolve binaries with a chirp mass of $M_c=1.187\usk M_\odot$, 
which is compatible with the very precise measurement result of  
$M_c = 1.186^{+0.001}_{-0.001} \usk M_\odot$ for GW event GW170817 
\cite{LVC:BNSSourceProp:2019}.
We consider the equal-mass case and one unequal-mass system with mass ratio
$q=0.9$. 
The NSs in our models are non-spinning, i.e. we use irrotational
initial data. However, note that spin can have an impact on many aspects
discussed here, as demonstrated, e.g., in \cite{Kastaun:2017, Chaurasia:2020,
Dietrich:2016:044045, East:2016:024011} for different systems.
The characteristic properties 
of our models are listed in \Tref{tab:models}. 

As initial data EOS, we use the lowest temperature of 
$0.1 \usk\mega\electronvolt$ that is available in the SFHO EOS.
The initial electron fraction for a given density is set according to 
$\beta$-equilibrium.
This approximation to the zero-temperature EOS breaks 
down at very low densities because the thermal pressure contribution 
becomes important and stays constant once it is dominated by the 
photon gas. To avoid technical problems determining the NS surface, 
we therefore replace the finite temperature table at densities  
below $1.4\times 10^7 \usk\gram\per\centi\meter\cubed$ by a matching
polytropic EOS (with adiabatic exponent $1.58$).

The effective tidal deformability $\tilde{\Lambda}$, which encodes the impact 
of tidal effects on the gravitational waveform during coalescence, is almost 
identical for the two mass ratios (in contrast to the individual deformabilities). 
The value of $\tilde{\Lambda}$
is compatible with upper limits inferred for GW170817 under the assumption of small 
NS spins \cite{LVC:BNSDetection,LVC:BNSSourceProp:2019,LVC:EOSPaper:2018,SoumiDe:2018}.
The statistical interpretation of the lower confidence bounds given in 
\cite{LVC:BNSSourceProp:2019,LVC:EOSPaper:2018,SoumiDe:2018} is called into question
\cite{Kastaun:2019}, but in any case $\tilde{\Lambda}$ is well above those limits
for our models. 
We also note that the Bayesian model selection study \cite{LVC:EOSModelSel:2020} does not rule
out even the zero tidal deformability case.

We note that our model is not compatible with the lower limit $\tilde{\Lambda}>450$ 
derived in \cite{Radice_2018} using inferred ejecta mass requirements for kilonova 
observation AT2017gfo. However, this value is based on an invalid assumption about 
the relation between disk mass and effective tidal deformability. A first counterexample
was found in \cite{Endrizzi:2018} and a systematic investigation \cite{Kiuchi:2019:876L}
provided more. Revised fitting formulas presented in \cite{Nedora:2020:11110}
exhibit large residuals and the search for robust analytic modeling of ejecta 
masses is an ongoing effort.   
In any case, the disk and ejecta mass is computed in our
simulations and will be compared to values inferred from the kilonova directly.

In order to compute BNS systems in quasi-circular orbit,  
we employ the \texttt{LORENE} code \cite{Gourgoulhon:2001:64029}. 
Since we are mainly interested in the qualitative post-merger behavior,
we take no steps to reduce the residual eccentricity inherent in the 
quasi-stationary approximation, and we chose an initial separation 
that corresponds to no more than 6 full orbital cycles before merger.

\begin{table}
\caption{Initial data parameters: $M_\mathrm{B}$ denotes the 
  total baryonic mass of the binary, $M_c$ the chirp mass, $M_1$ and 
  $M_2$ the gravitational masses of the stars, $q=M_2/M_1$ the mass 
  ratio, $\Lambda_1$ and $\Lambda_2$ the dimensionless tidal 
  deformability of the stars, and $\tilde{\Lambda}$ the effective 
  tidal deformability.}
\begin{ruledtabular}
\begin{tabular}{lcc}
Model&
\texttt{Q10}&
\texttt{Q09}
\\\hline 
$M_\mathrm{B}\,[M_\odot]$&
$3.001$&
$3.008$
\\
$M_c\,[M_\odot]$&
$1.187$&
$1.187$
\\
$M_1\,[M_\odot]$&
$1.364$&
$1.438$
\\
$M_2\,[M_\odot]$&
$1.364$&
$1.294$
\\
$q$&
$1.0$&
$0.9$
\\
$\Lambda_1$&
$396$&
$280$
\\
$\Lambda_2$&
$396$&
$551$
\\
$\tilde{\Lambda}$&
$396$&
$396$
\end{tabular}
\end{ruledtabular}
\label{tab:models}
\end{table}
 
Using the same EOS as for the initial data, we computed the baryonic mass for sequences of 
NSs rotating uniformly with rate at the mass shedding limit (using the \texttt{RNS} code 
\cite{Stergioulas:1995:306}). We find that the maximum baryonic mass for a uniformly rotating NS is 
$2.86 \usk M_\odot$. Based on comparisons in \cite{Kaplan:2014:19,Kastaun:2016}, we do not expect 
thermal contributions in the heated merger remnant to significantly increase this maximum. 
The total baryonic mass of our BNS models is well above the maximum allowed for uniformly
rotating models. Even allowing for atypically large mass ejection of $0.1\usk M_\odot$,
the remnant is therefore hypermassive, i.e., it requires non-uniform rotation to delay 
collapse. We therefore expect BH formation within tens of ms after merger.

\subsection{Coordinate Systems}
\label{sec:gauge}

The standard 1+log and gamma-driver gauge conditions used during evolution are 
well suited to prevent catastrophic gauge pathologies,
but they are not designed to recover axisymmetric coordinates when the spacetime
approaches a mostly axisymmetric stationary phase. The coordinate system present
after merger depends not on the final mass distribution, but on the whole history of the 
evolution. Therefore, one cannot rely on the coordinate-dependent quantities,
e.g. multipole moments expressed in coordinates, to measure any deviations
from axisymmetry. 

In \cite{Kastaun:2015:064027}, we developed a postprocessing procedure to obtain a
well defined coordinate system in the equatorial plane with the following 
properties
\begin{enumerate*}
\item The radial coordinate is the proper distance to the origin along radial
coordinate lines
\item The angular coordinate is based on proper distance along arcs of constant radial
coordinate
\item On average, the radial coordinate lines are orthogonal on the angular ones,
thus minimizing twisting.
\end{enumerate*} 
If the spacetime is indeed axisymmetric (with axis orthogonal to the equatorial 
plane of the simulation coordinates) then so are the new coordinates.

In this work, we also want to study the 3D structure of the remnant. We therefore
need to extend the coordinate system above from the equatorial plane. However,
the metric was not saved in 3D in our simulations, which precludes a generalization 
in the same spirit. Instead, we use an ad-hoc construction as follows.
First, we apply the same coordinate transformation as within the equatorial 
plane to all planes with constant $z$-coordinate. 
Using the metric saved along the $z$-axis during the simulation, we transform the 
$z$-coordinate as $z \to z'(z)$ such that on the $z$-axis, the new $z$-coordinate 
is the proper distance to the equatorial plane along the axis. Away from the axis,
the new $z$-coordinate is only an approximation to the proper distance to the
equatorial plane.

The resulting 3D coordinate system allows to judge axisymmetry in the equatorial 
plane, and it allows to assess oblateness since distances along the $z$-axis and 
within the equatorial plane are exact proper distances. In the meridional planes
a coordinate circle might still show some deviations from a proper sphere,
except on $z$-axis and the equator. In the rest of this work, we refer to this 
coordinate system as postprocessing coordinates to distinguish from simulation 
coordinates.

From previous experience \cite{Kastaun:2016,Kastaun:2017,Ciolfi:2017:063016}, 
we expect that the remnant is changing only slowly 
when viewed in a coordinate system rotating with a certain angular velocity,
which is also changing slowly (also compare the animations provided 
in the supplemental material of \cite{Kastaun:2017}).
In other words, we expect an approximate helical
Killing vector. 

To extract the rotating pattern of mass distribution and  
velocity field, we construct corotating coordinates as follows. First,
we perform a Fourier decomposition with respect to $\phi$ (in postprocessing
coordinates) in the equatorial plane. We then compute a density-weighted
average to get the phase of the dominant $m=2$ density deformation as function 
of time. We further apply a smoothing by convolution with a $2\usk\milli\second$
long Hanning window function to suppress high frequency contributions.
We then apply the opposite rotation to the 3D postprocessing 
coordinates at each time to obtain postprocessing coordinates corotating
with the main deformation pattern.

\subsection{Diagnostic Measures}
\label{sec:diag}

In order to extract gravitational waves, we decompose the Weyl-scalar $\Psi_4$
into spin-weighted spherical harmonics, considering all multipole coefficients 
up to $l=4$. The strain is computed by time integrating using the 
fixed-frequency integration \cite{Reisswig:2011:195015} method with a
low-frequency cutoff at $500\usk\hertz$. The fluxes of energy and angular 
momentum are also computed using multipole components up to $l=4$.
We use a fixed extraction radius 
$R_\mathrm{ex} = 916 \usk\kilo\meter$, close to the outer boundary of the 
computational domain. 
We do not extrapolate the signal to infinity 
as we expect the finite resolution error to dominate the error due to 
finite extraction radius. 

To describe the distribution of matter, we use the baryonic mass density $\rho$, 
defined as baryon number density in the fluid rest frame times an arbitrary mass 
constant (in this work $1.66\times 10^{-27} \usk\kilo\gram$). Baryon number conservation 
implies a conserved current $u^\mu \rho$, with $u$ being the fluid 4-velocity.
The total baryonic mass within a volume can only change by matter leaving 
the boundary, and is given by
\begin{align}
M_\mathrm{B} &= \int_V W \rho \,\mathrm{d}V, & 
\mathrm{d}V &= \sqrt{\gamma} \,\mathrm{d}^3x
\end{align} 
where $W$ is the Lorentz factor of the fluid with respect to Eulerian observers, 
$\mathrm{d}V$ is the proper 3-volume element and  
$\gamma$ is the determinant of the 3-metric.
On the numerical level, the baryonic mass definition is 
complicated by the use of an artificial atmosphere. Our numerical volume 
integrals of baryonic mass exclude any grid cell set to atmosphere. 

To obtain the mass of dynamically ejected matter, we compute the 
time-integrated flux of unbound matter through several coordinate spheres 
with radii between $73$--$916\usk\kilo\meter$. For each sphere, we then 
add the volume integral of residual unbound matter still present within the same
sphere at the end of the simulation. 
Matter is considered unbound according to the 
geodesic criterion $u_t<-1$, where $u$ is the 4-velocity, and the artificial
atmosphere is excluded. This criterion assumes force-free ejecta and therefore
becomes more accurate at larger radii.

The combined measure for the ejecta mass alleviates drawbacks of using flux or 
volume integrals only. 
When using only the flux through an extraction sphere, one is either restricted 
to small extraction radii to ensure that all ejecta are accounted for, or 
forced to evolve the system long enough to allow all ejecta to reach the extraction 
radius. When using only volume integrals, they have to be computed before significant 
amounts of ejecta leave the computational domain. Such integrals then include matter at small 
radii where the geodesic criterion is unreliable, and miss ejecta that become unbound 
later. The combined measure allows meaningful comparison over a larger range of 
extraction radii. For the cases at hand, we find negligible differences for radii 
$\gtrsim 400\usk\kilo\meter$, and use the outermost radius $916\usk\kilo\meter$ 
for quoting ejecta masses.

We employ a similar approach for estimating the escape velocity. 
In a stationary spacetime, the velocity that a fluid element would reach 
after escaping the system on a geodesic trajectory is $v^2_\infty = 1-u_t^{-2}$.
Again, we consider both the ejecta leaving the system through a spherical extraction 
surface during the simulation, and the unbound matter still within the domain at the 
end. Both contributions are filled into a mass-weighted histogram of the escape
velocity. This way, we account for the fastest components via the flux as well as the 
slowest ones via the unbound matter at final time.

For technical reasons, we first combine all ejecta at a given time within thin 
spherical shells $V_s$ with radius $R_s$ and thickness $\delta R_s$.
For those, we compute the volume integrals
\begin{align}
W_s &= \frac{1}{M_s} \int_{V_s} u_t W \rho_u \,\mathrm{d}V ,&
M_s &=\int_{V_s} W \rho_u \,\mathrm{d}V
\end{align}
where $\rho_u$ is the density of unbound matter in the fluid rest frame. 
From the above measures, we attribute an average escape velocity
$v^2_s = 1-W_s^{-2}$ to each shell. 
To compute the above integrals at each time for each radius, we employ a 
simple and robust technical implementation based on creating histograms 
of all numerical grid cells, binned by radius, and weighted by the integrands.

Another quantity relevant for our study is the ADM mass of the system.
This measure is formally defined for the whole spacetime, as there is no
locally conserved energy in GR. It can be expressed either via surface
integrals at infinity, or 3-volume integrals over a spacelike hypersurface 
of a given foliation of spacetime. When restricting either formulation
to sufficiently large but finite region, 
such that the outer boundary lies in the weak field regime, 
then the ADM mass is not constant but changes by the amount of energy 
carried away by GW. We thus compute a time-dependent ADM mass from the 
ADM mass of the initial data minus the integrated GW energy flux.

As a heuristic measure of energy distribution, we monitor the integrand 
of the ADM mass volume integral as well.
However, we stress that this
is not gauge invariant as it depends on the chosen foliation of spacetime.
Our motivation is to split the total ADM energy into contributions
of the remnant NS and the surrounding disk. 

We also need to consider the gravitational radiation still inside the 
computational domain. As a practical measure, we use the following:
considering the region between $R_0 < r < R_\mathrm{ex}$, we define 
a GW energy at time $t$ as the integrated GW flux through $R_\mathrm{ex}$
over the time interval $(t, t+R_\mathrm{ex} - R_0)$. 
We compute this measure for $R_0$ as low as 
$100\usk\kilo\meter$ (around the wavelength of a $3\usk\kilo\hertz$ signal).
In other words, we associate an energy loss of the remnant at a given time 
with the GW luminosity at a large extraction radius at the time when the 
radiation from radius $R_0$ has reached the extraction radius.
This can only provide a qualitative picture, as the measure is build on  
concepts valid in the weak field limit / wave zone. 

For the angular momentum, we use the volume integral formulation 
of the ADM angular momentum, similarly to the ADM mass above. 
As for the mass, we compute the GW angular momentum loss, at the same
extraction radius.
For axisymmetric spacetimes, another angular momentum definition is given by 
Komar. 
Since the system approaches a roughly axisymmetric state after merger, it
makes sense to use the Komar angular momentum. 
We note that the Komar measure is more closely related to the fluid in the 
sense that there are no contributions of vacuum, horizons aside, and, 
consequently, no contributions of GW radiation present in the system. 
For exact definitions of ADM and Komar quantities, we 
refer to \cite{Gourgoulhon:2012:846}.

The post-merger evolution is not exactly axisymmetric, the postprocessing 
coordinates are not available during the simulation, and we avoid storing all 
metric quantities as 3D data. Therefore, we use an approximation to the Komar 
angular momentum that is obtained by integrating the $\phi$-component (in 
simulation coordinates) of $S_\phi$, the evolved quasi-conserved momentum 
density. This is trivial to compute during the simulation, but becomes exact 
only if the $\phi$-coordinate is a Killing vector field. 

Besides volume integrals over the full domain, we are interested in the 
radial distribution of the integrands. For this, we employ a numerical 
method developed in a previous study \cite{Kastaun:2016}. This method 
allows an efficient computation of volume integrals 
\begin{enumerate*}[label=(\roman*)]
\item within spheres of constant coordinate radii as function of radius, and
\item within regions above given mass densities $\rho$ as function of $\rho$.
\end{enumerate*}
It works by adding the integrand (including the volume element) in each 
numerical cell  
into onedimensional 
histograms binned in terms of coordinate radius and density, respectively.
The volume integrals can then be obtained during post-processing simply
by cumulative summation over the bins.
Using this method, we integrate the
\begin{enumerate*}[label=(\roman*)]
\item proper volume
\item baryonic mass
\item ADM mass
\item ADM angular momentum
\item Komar angular momentum approximation
\item estimated mass of unbound matter
\end{enumerate*}.

We note that the 3-dimensional isodensity surfaces in 4-dimensional spacetime 
are gauge-independent. The corresponding volume integrals within a time slice 
depend only on the time slicing but not on the spatial gauge.
In contrast, integrals over spheres of constant coordinate radius depend
on the spatial coordinates. In order to reduce this dependency, we parametrize
the spheres by the enclosed proper volume. The only remaining gauge ambiguity 
(beside the time slicing) is given by the shapes of the coordinate spheres, but 
not their overall extent.

Similarly, we also parametrize the integrals 
within isodensity surfaces by the enclosed proper volume. Within the remnant NS,
where the mass distribution is roughly spherical, the two methods of defining
radial mass distribution should roughly agree. This is not the case for
the torus-shaped disk.
For convenience, we will sometimes 
express proper volumes in terms of the radius of Euclidean spheres with same 
volume (``volumetric radius'', $R_V$). 

Based on the above integrals, we can define a measure for the compactness
of any volume as the baryonic mass divided by the volumetric radius.
The compactness of isodensity surfaces as function of volumetric radius a
has a maximum. We refer to the region within this maximum-compactness surface
as the ``bulk''. We use the bulk definition to divide the matter distribution
after merger into a remnant and a disk, but we stress that this is somewhat 
arbitrary as there is a smooth transition.

\section{Results}
\label{sec:implic}

\subsection{Overall dynamics}  
\label{sec:dynamics}

\begin{figure}
\includegraphics[width=0.95\columnwidth]{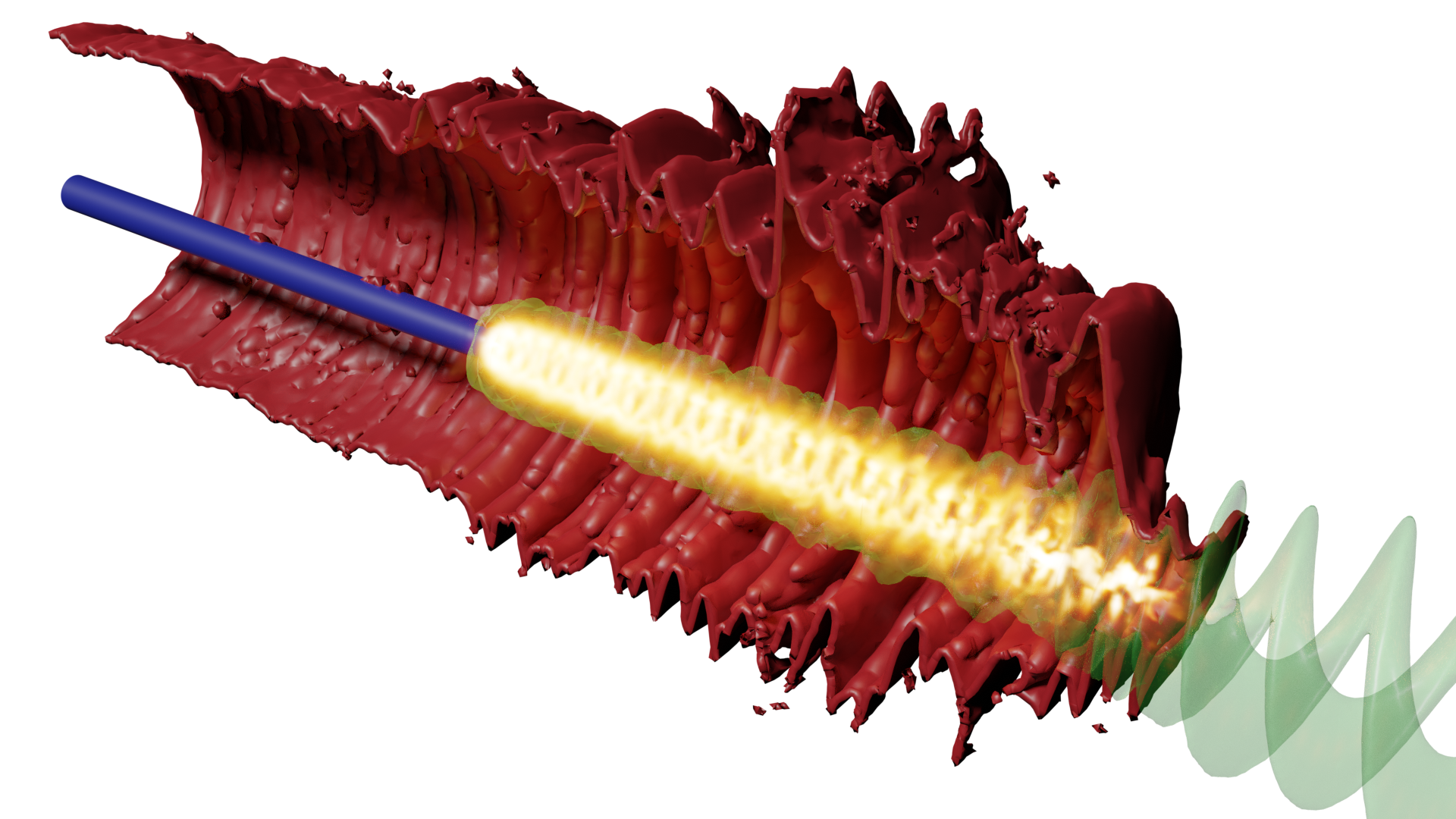} 
\caption{Overview of the evolution phases for the $q=0.9$ case. Time runs 
from lower right to upper left, while the
other two dimensions correspond to the orbital plane. The transparent green 
surface corresponds to a fixed density of 
$5 \times 10^{13}\usk\gram\per\centi\meter\cubed$, highlighting the evolution
of the merged NS and the coalescing NS shortly before merger. The solid red 
surface corresponds to a density of $10^{11}\usk\gram\per\centi\meter\cubed$,
as a proxy for the denser parts of the disk. To avoid occlusion, one half-plane 
was cut away. The blue surface marks the apparent horizon extracted during the 
simulation. To give an impression of the thermal evolution inside the remnant,
we rendered hot, dense regions as light emitters (without absorption).
The time coordinate was compressed by a factor $0.05$ with respect to the spatial 
coordinates in geometric units, such that light cones would appear almost 
orthogonal to the world tube of the remnant.
}\label{fig:remnant_xyt}
\end{figure}

In the following, we provide a broad overview on the merger outcome.
Key quantities are summarized in \Tref{tab:outcome}.
Qualitatively, both cases are very similar. For the example of the 
$q=0.9$ case, we visualize the evolution timeline in \Fref{fig:remnant_xyt}.
The coalescing NS merge into a hypermassive NS (HMNS) which collapses to a 
BH after a delay on the order of ${\approx}10\usk\milli\second$. 
The HMNS is embedded in a 
massive debris disk created during merger. The disk is strongly perturbed 
by interactions with the remnant, and settles to a more stationary state 
shortly after the BH is formed.

\begin{figure}
\includegraphics[width=0.95\columnwidth]{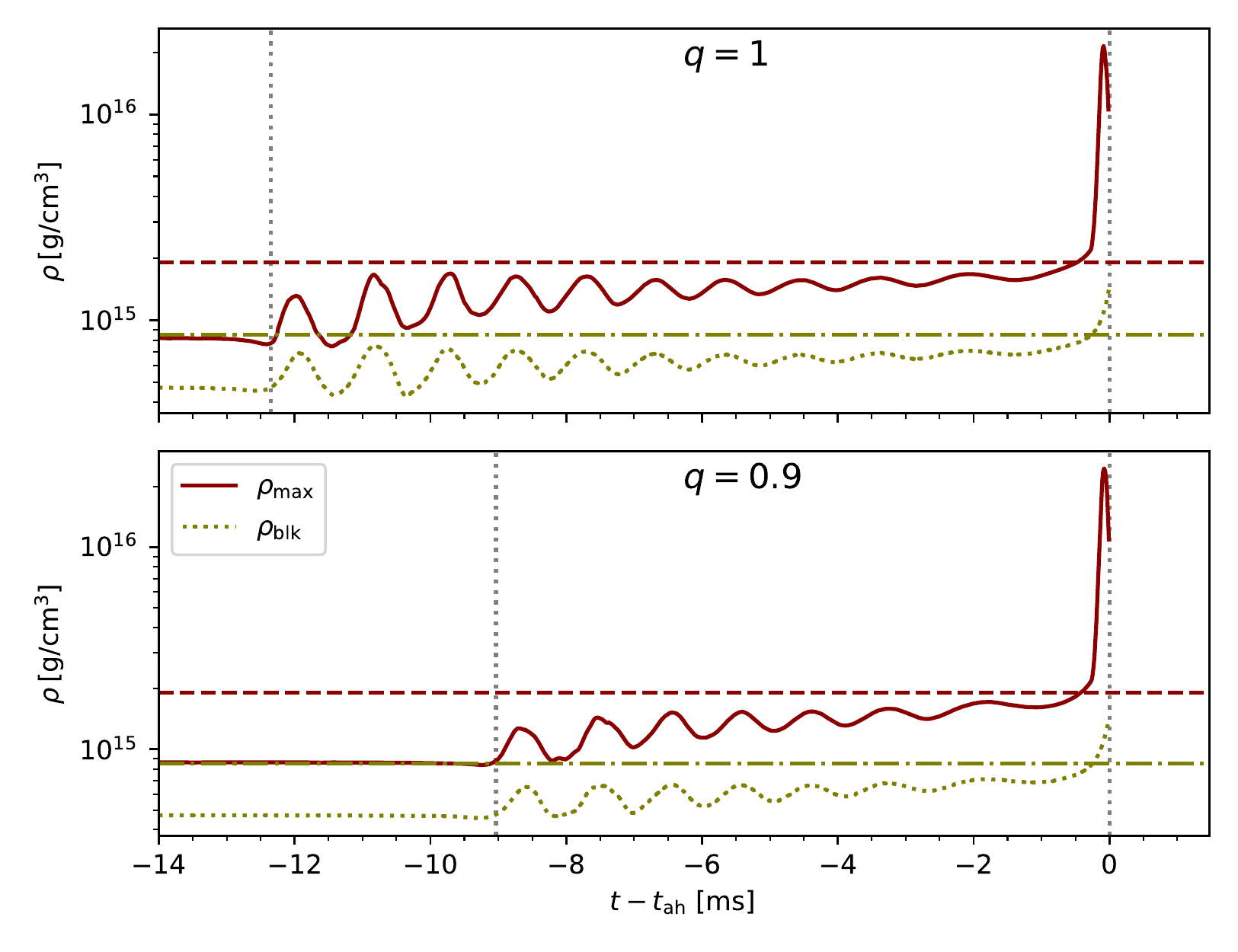} 
\caption{Post-merger evolution of mass density in the remnant
for the equal mass case (top panel) and unequal mass case (bottom panel).
The solid curve shows the maximum baryonic mass density in the fluid frame.
The dotted curve shows an average density given by bulk mass per
bulk volume (see text). The time refers to coordinate time.
For comparison, the horizontal lines show the 
central density (dashed) and average bulk density (dash-dotted)
of the maximum-mass TOV solution. The vertical lines mark the time of 
merger and formation of an apparent horizon. 
}\label{fig:dens_evol}
\end{figure}

Quantitatively, we observe some differences between the two mass ratios.
Most notably, the 
collapse delay is about 0.25\% shorter for the unequal mass case. This can be seen 
in \Fref{fig:dens_evol} showing the evolution of the remnant density.
We stress that in general, the delay time  
for a given mass is very sensitive to numerical 
errors, because the system is at the verge of collapse (we will discuss 
the evolution leading to collapse in later sections).
However, since both simulations employ the same resolution, grid setup, and 
numerial method, we expect the difference of the delays to be more robust 
than the absolute values. 

For a remnant close to collapse, one can expect that small changes
of the initial parameters, mainly the total mass, should lead to large changes
of the delay. This is however not a drawback. Any observational constraint
on the delay 
translates into a stronger constraint on the total 
mass. In this context, the relevant numerical uncertainty is not the (large) error 
of the delay for a given mass, but the (smaller) error in the total mass that leads to 
a given collapse delay.

We emphasize that our non-magnetized simulations exclude
the possibility of effective magnetic viscosity due to small-scale
magnetic field amplification. 
Such effects might reduce the collapse delay further. Since it is difficult to
predict the impact of mass ratio on magnetic field amplification,
we also cannot exclude an impact on the relation between collapse delays
and mass ratio.
For further discussion, see \cite{Kiuchi:2018:124039}
and the references therein.

\begin{figure}
\includegraphics[width=0.95\columnwidth]{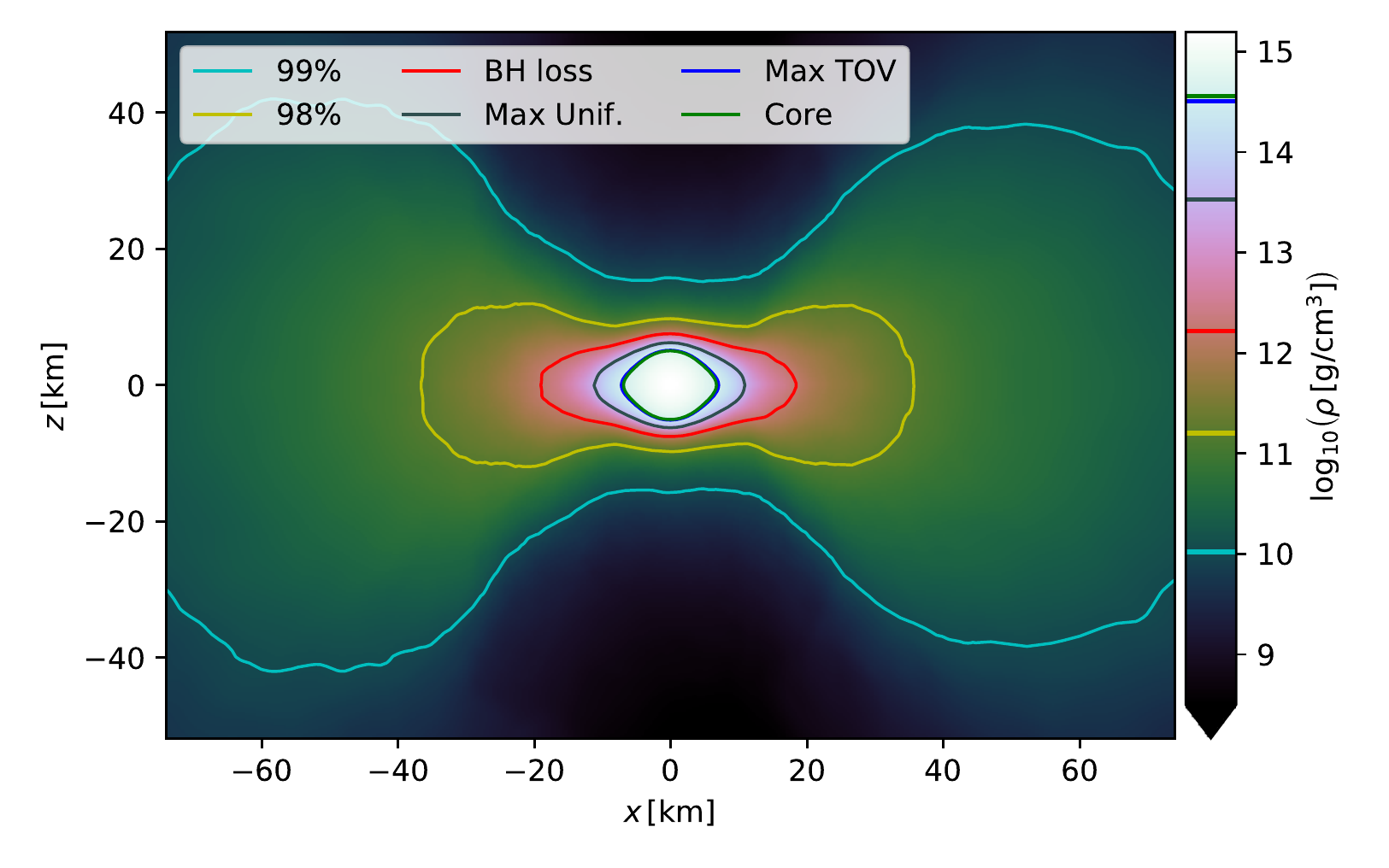}\\
\caption{Mass distribution in the meridional plane 
$2\usk\milli\second$ before apparent 
horizon formation, averaged over a time window $\pm 1 \usk\milli\second$. 
The color scale shows the baryonic mass density. The contours mark densities
for which the corresponding isodensity surfaces contain selected mass
fractions. We show the contours for $99\%$, $98\%$, the mass swallowed by
the BH within $1 \usk\milli\second$ after formation, the maximum mass of 
uniformly rotating and nonrotating NS, and the (nearly identical) mass of 
the nonrotating NS best approximating the core (see text).
}\label{fig:ns_disk_xz}
\end{figure}

\begin{figure}
\includegraphics[width=0.95\columnwidth]{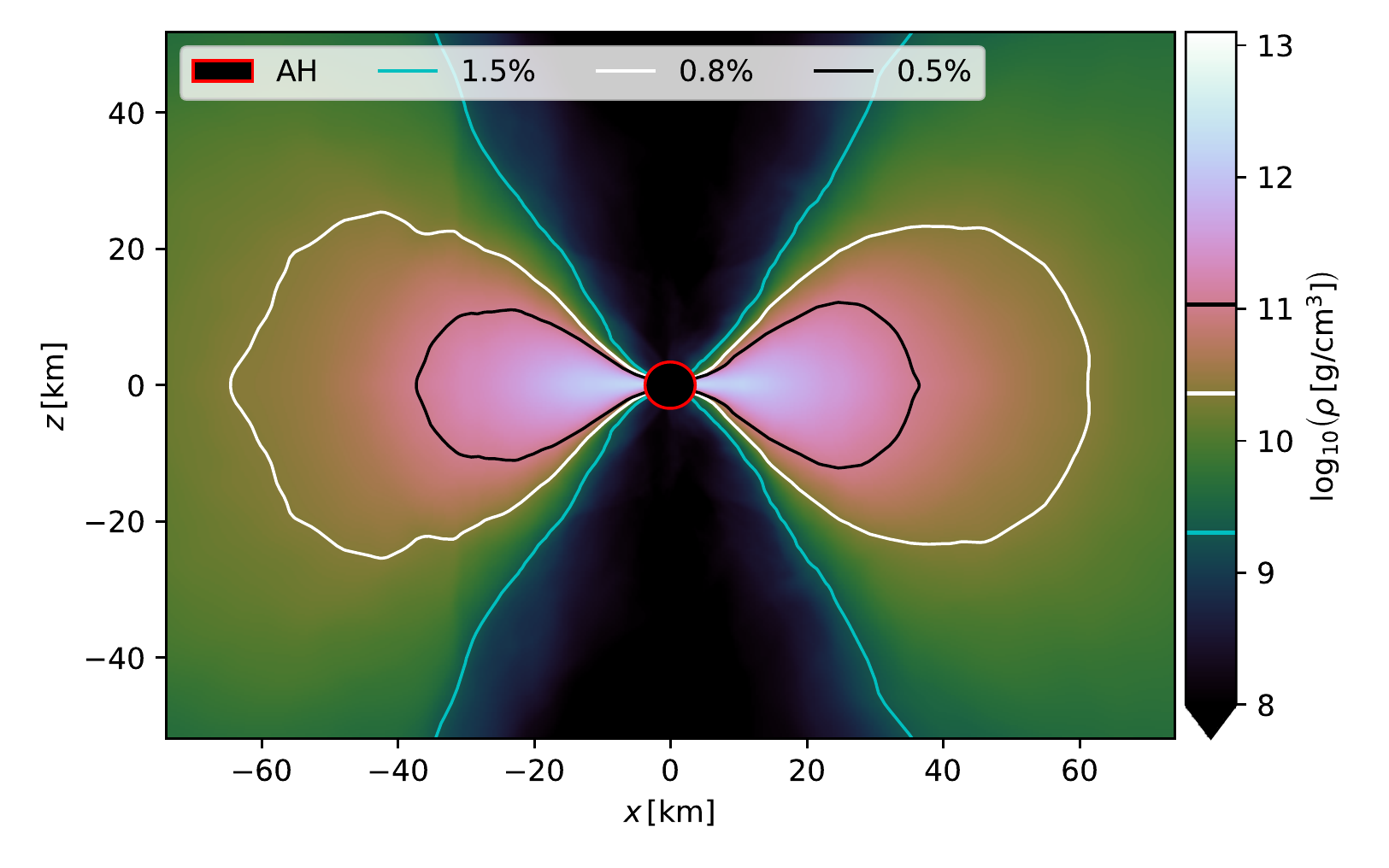}\\ 
\caption{Like \Fref{fig:ns_disk_xz}, but showing the time
$2\usk\milli\second$ after apparent horizon formation. 
The density contours contain selected fractions of initial total baryonic 
mass (excluding the BH interior).
}\label{fig:bh_disk_xz}
\end{figure}

The HMNS is smoothly embedded within a debris disk. The structure of the disk 
is shown in \Fref{fig:ns_disk_xz} for the example of the $q=0.9$ case.
The innermost part of the disk falls into the BH after the remnant 
collapses. After collapse, the remaining 
disk mass is around $0.05\usk M_\odot$ (see \Tref{tab:outcome}).
The disk contains enough matter to supply a wind that could explain the red 
component of the kilonova AT2017gfo observed after GW170817, with an inferred
mass ${\approx}0.04\usk M_\odot$ \cite{Kasen:2017:551}. 
It would, however, require an effective mechanism in order to 
expel around 80\% of the disk. 
The structure of the disk after BH formation 
is shown in \Fref{fig:bh_disk_xz} for the $q=0.9$ example.

As a general trend, we expect that for fixed EOS and mass ratio, systems with lower total
mass possess a more massive disk after merger (compare, for example, \cite{Radice_2018}). 
We will further show that in our cases, some matter is migrating from the 
remnant into the disk. This was already observed for different models in previous 
studies \cite{Kastaun:2016,Endrizzi:2018}. Since the remnant lifetime
also increases with decreasing mass, the final disk mass should depend even more
strongly on the total mass. 
Conversely, we expect a more massive disk for a system with the same total 
mass, but obeying a different EOS for which the maximum NS mass is slightly 
larger.

\begin{table}
\caption{Key parameters of merger outcome. $M_\mathrm{BH}$ and 
$J_\mathrm{BH}$ are black hole mass and angular momentum 
$5.0\usk\milli\second$ after formation. 
$F_\mathrm{BH}$ is the GW frequency from quasi-normal-mode ringdown
for the given mass and spin. 
$M_\mathrm{blk}$ and $R_\mathrm{blk}$ are bulk mass and bulk 
volumetric radius, extracted 
$1.0\usk\milli\second$ before collapse.
Rows $\nu_\mathrm{cnt}$ and $\nu_\mathrm{max}$ denote the 
remnants central and maximum rotation rates computed 
$1.0\usk\milli\second$ before collapse.
$f_\mathrm{merge}$ is the gravitational wave instantaneous frequency 
at the time of merger, $f_\mathrm{pm}$ is the frequency of the 
maximum in the post-merger part of the gravitational wave power 
spectrum. 
Row $M_\mathrm{disk}$ provides the baryonic mass outside the 
apparent horizon, excluding unbound matter, at time 
$6.0\usk\milli\second$ after collapse.
$M_\mathrm{ej}$ is the estimate for the total
mass of dynamically ejected matter,  $v_\infty$ refers to 
the median, 5th and 95th percentiles of the mass-weighted 
escape-velocity distribution of ejected matter.}
\begin{ruledtabular}
\begin{tabular}{lcc}
Model&
\texttt{Q10}&
\texttt{Q09}
\\\hline 
$M_\mathrm{BH}\,[M_\odot]$&
$2.55$&
$2.57$
\\
$J_\mathrm{BH}/M^2_\mathrm{BH}$&
$0.66$&
$0.67$
\\
$F_\mathrm{BH}\, [\mathrm{kHz}]$&
$6.56$&
$6.52$
\\
$M_\mathrm{blk}\,[M_\odot]$&
$2.56$&
$2.59$
\\
$M_\mathrm{blk}/R_\mathrm{blk}$&
$0.31$&
$0.31$
\\
$\nu_\mathrm{cnt}\, [\mathrm{kHz}]$&
$0.96$&
$0.87$
\\
$\nu_\mathrm{max}\, [\mathrm{kHz}]$&
$1.76$&
$1.71$
\\
$f_\mathrm{merge}\, [\mathrm{kHz}]$&
$1.94$&
$1.94$
\\
$f_\mathrm{pm}\, [\mathrm{kHz}]$&
$3.38$&
$3.35$
\\
$M_\mathrm{disk}\,[10^{-2} \, M_\odot]$&
$5.5$&
$4.6$
\\
$M_\mathrm{ej}\, [10^{-2} \, M_\odot]$&
$1.7$&
$0.8$
\\
$v_\infty\, [c]$&
$0.16^{+0.08}_{-0.11}$&
$0.14^{+0.09}_{-0.08}$
\end{tabular}
\end{ruledtabular}
\label{tab:outcome}
\end{table}

\begin{figure}
\includegraphics[width=0.95\columnwidth]{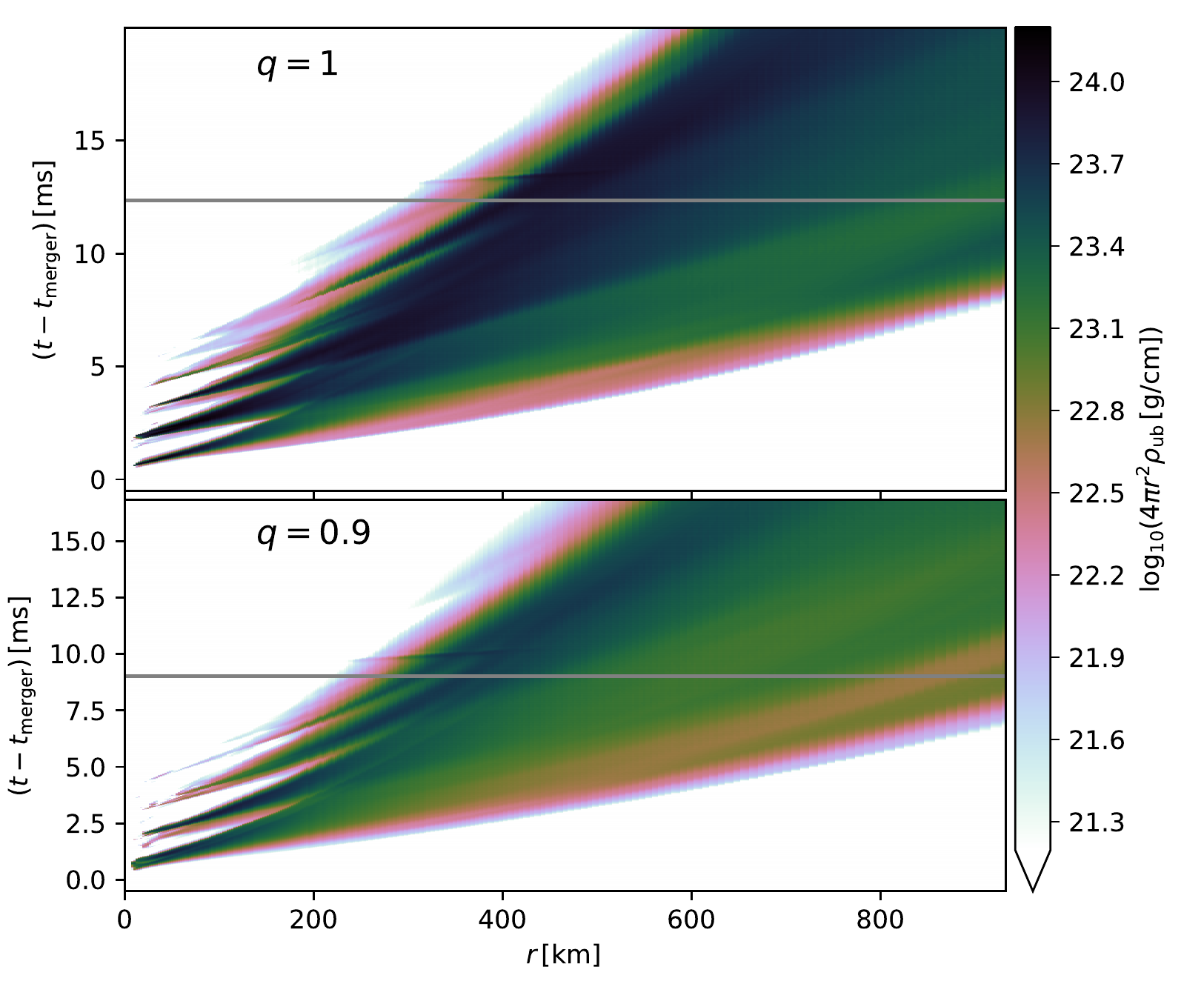}\\ 
\caption{Radial distribution of unbound matter versus time after merger. 
The color scale corresponds to unbound mass per radial distance, where matter
is considered unbound according to the geodesic criterion.
The horizontal lines mark the time of BH formation.
}\label{fig:ejecta_rc}
\end{figure}

We observe significant dynamical mass ejection during merger and during the 
remnant lifetime. 
As shown in \Fref{fig:ejecta_rc}, several independent mass ejections are launched, 
mostly from radii ${\lesssim} 100 \usk\kilo\meter$. The individual ejected components 
merge, because of their velocity dispersion, and leave the system as 
a single ejecta component.
It seems that the pressure waves injected into the disk (see also \Fref{fig:remnant_xyt}) 
by the HMNS also result in matter ejection from the disk. 

The ejecta mass as extracted from the numerical results is given in 
\Tref{tab:outcome}. We stress that in general, ejecta masses extracted from 
numerical simulations are affected comparably strong by the numerical errors.  
Convergence tests presented in \cite{Ciolfi:2017:063016} for simulations of a
long-lived remnant using a very similar numerical setup found a finite resolution 
error of the dynamical ejecta mass around $20\%$. 
In our case, an additional --- and likely dominant --- source of uncertainty is the 
dependence of dynamical ejecta mass on the lifetime of the remnant. The latter 
can be extremely sensitive to errors if the remnant is on the verge of collapse. 
It is therefore difficult to estimate the error without expensive tests with much 
higher resolutions, but we suspect that the error could easily reach a factor two.

Comparing to other results in the literature, we find a disk mass that is an outlier 
to the phenomenological fit of disk mass in terms of effective tidal 
deformability that was proposed in \cite{Radice_2018}. Other counterexamples 
were already found in \cite{Endrizzi:2018,Kiuchi:2019:876L}.
We also note that \cite{Nedora:2021:98} performed a simulation that corresponds 
almost exactly to our $q=1$ setup, except that it includes neutrino radiation. 
They quote a lower ejecta mass $(2.8\pm 0.7) \times 10^{-3} \usk M_\odot$
as well as a lower disk mass $(1.9 \pm 0.7)\times 10^{-2} \usk M_\odot$,
but also find a remnant lifetime that is around $3$ times shorter than for our
simulations. 

Our results depicted in \Fref{fig:ejecta_rc} show that mass is continuously 
ejected during the HMNS lifetime. The differences in lifetime could thus account 
for the tension regarding the ejecta masses. Similarly, it might account as least 
partially for the lower disk mass. The difference in lifetime could be due to finite 
resolution errors alone, but it is also possible that the inclusion of neutrino 
transport has an influence on HMNS close to collapse. In any case, the result that
increasing the lifetime of a HMNS can lead to larger disk and ejecta masses
suggests that fitting these quantities in terms of the binary parameters as in 
\cite{Radice_2018,Nedora:2020:11110} is challenging for the parameter ranges 
resulting in short-lived HMNS. We propose to include the lifetime as an unknown 
in the fit, not just because of the sensitivity with regard to total mass and 
numerical errors, but also because the lifetime may be affected by physical effects
such as magnetic viscosity that are essentially unknown.

The spectral evolution of the kilonova depends strongly on the ejecta velocity.
\Tref{tab:outcome} reports the median of the velocity distribution extracted 
from our simulations as described in \Sref{sec:diag}, together with 5th and 95th 
percentiles.
The values refer to the outermost extraction radius, but we also compared 
smaller ones. We find that the median and 5th percentile are stable outside 
$400\usk\kilo\meter$, whereas the 95th percentile continuously decreases and 
should be considered as unreliable. Given that the fastest components are those 
running into the artificial atmosphere, the deceleration is likely unphysical.
Another source of uncertainty is that an earlier collapse of the HMNS would 
result in less ejected mass, but faster median velocity, because ejecta launched 
at later times tend to be slower for the cases at hand.

The mass and velocity found in our numerical results are both about a factor 
two lower than the estimates inferred by \cite{Kasen:2017:551} 
for the blue component of the kilonova AT2017gfo.
However, because of the uncertainties discussed above, we cannot make a 
conclusive statement if the dynamical ejecta mass for the SFHO EOS is 
compatible with the observed kilonova.

Besides mass and velocity, kilonova models such as \cite{Kasen:2017:551} also  
predict a strong dependency on the composition of the ejecta. The initial electron fraction of
the neutron-rich ejecta is strongly affected by neutrino radiation (see, e.g., 
\cite{Martin:2018}). Since those are not included in our study, we refrain from 
discussing the ejecta composition, but note that once again the lifetime of the 
HMNS has a direct impact on an observable.

It should also be noted that disk evolution and ejecta might be sensitive to 
magnetic field effects, which are not included here.
For the example of a system with large initial magnetic field that was studied 
in \cite{Ciolfi:2019:023005}, the entire disk was driven to migrate outwards (but 
not necessarily ejected). On the other hand, a reduction of remnant lifetime 
by magnetic viscosity might result in less dynamical ejecta and a less 
massive disk.  

\subsection{Gravitational Waves}
\label{sec:gw}

In this section, we present the GW signals extracted from the 
simulations as described in Sec.~\ref{sec:diag}. We compare the dominant 
spherical harmonic $\ell=\vert m \vert = 2$ mode with predictions from theoretical 
waveform models, produce a hybrid waveform combining analytical inspiral 
data with the result from our numerical simulations, and quantify the initial 
eccentricity of our simulations through the GW frequency.

\begin{figure*}
\includegraphics[width=0.95\textwidth]{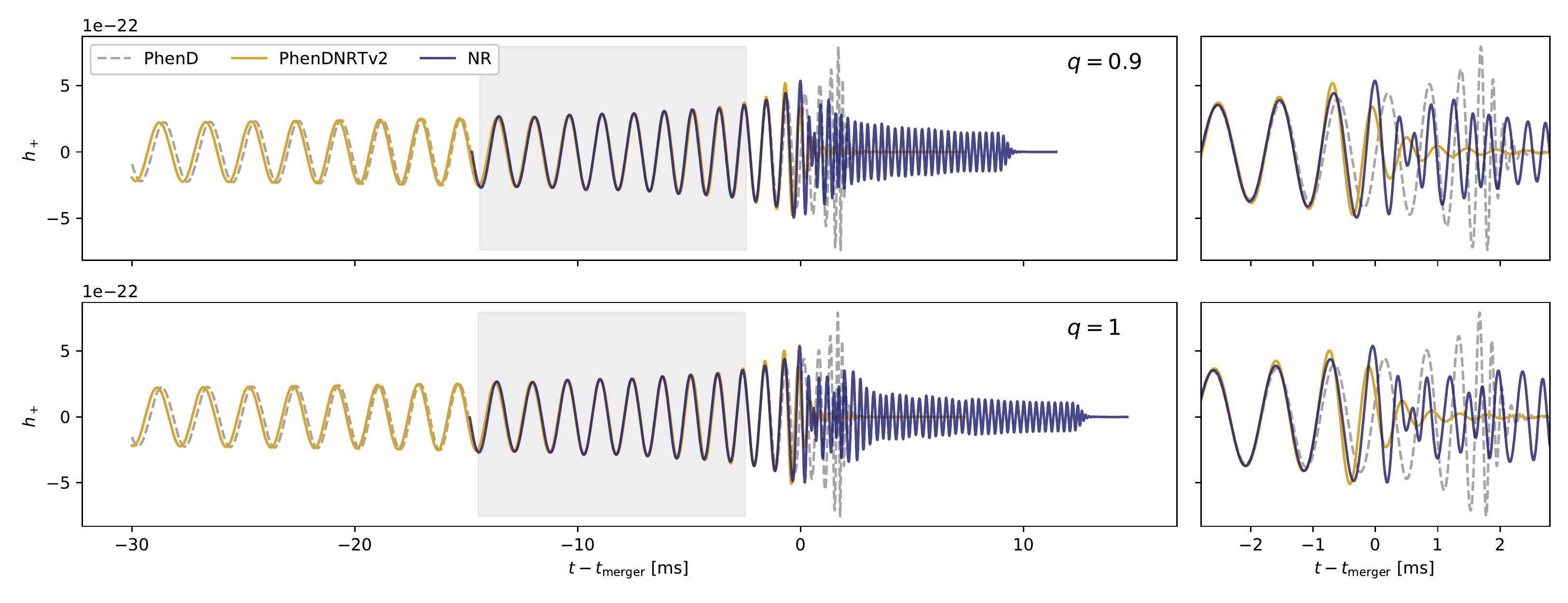} 
\caption{The GW signals extracted from our 
simulations (blue) as observed face-on at a distance of 
$40.7\,\textrm{Mpc}$. We extend the inspiral with the 
BNS model \texttt{IMRPhenomD\_NRTidalv2} 
\cite{Dietrich:2019kaq} (orange), aligned with the numerical data over the grey 
shaded region. For comparison, we also include the BBH model 
\texttt{IMRPhenomD}~\cite{Husa:2015iqa, Khan:2015jqa} (gray dashed lines).}
\label{fig:waveforms} 
\end{figure*}

\begin{figure}
\includegraphics[width=0.95\columnwidth]{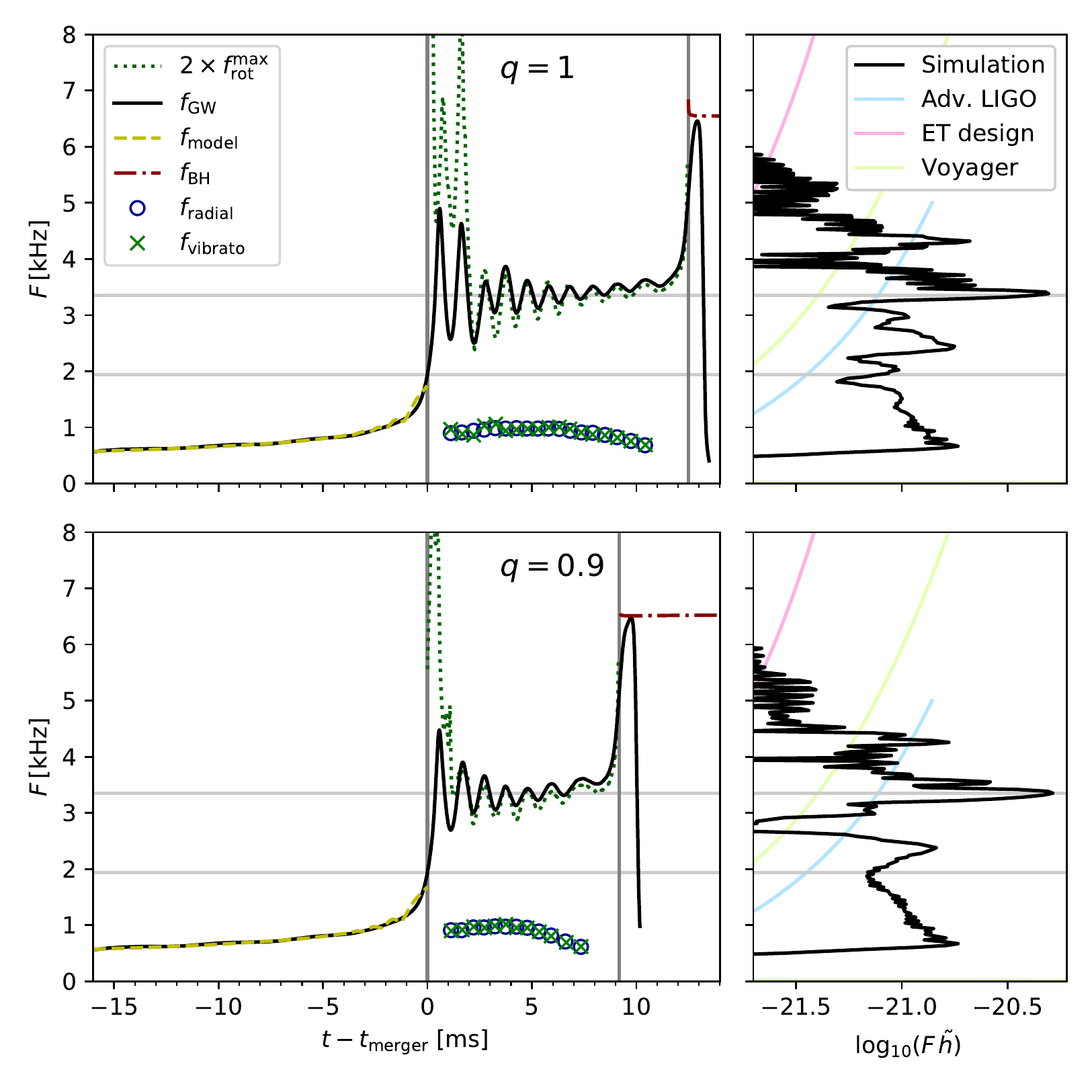} 
\caption{Left panels: time evolution of GW frequency 
$f_\mathrm{GW}$, maximum rotation rate $\nu_\mathrm{max}$, 
radial oscillation frequency $f_\mathrm{radial}$, 
and the rate of the modulation of the GW frequency, $f_\mathrm{vibrato}$.
Vertical lines mark time of merger and apparent horizon formation.
For comparison, we show the inspiral GW frequency $f_\mathrm{model}$ according 
to the \texttt{IMRPhenomD\_NRTidalv2} waveform model.
Further, we compute the $m=l=2$ BH quasinormal-mode frequency $f_\mathrm{BH}$  
from BH mass and angular momentum as found in the simulation at each time. 
Horizontal lines mark the frequency at merger (maximum GW amplitude) and 
the main peak of the spectrum.
Right panels: Power spectrum of the $l=m=2$ component of the GW signal,
at distance $40.7\usk\mega\mathrm{Pc}$, in terms of $F\tilde{h}(F)$,
where $\tilde{h}^2(F) = \tilde{h}_+^2(F) + \tilde{h}_\times^2(F)$.
For comparison we show the design sensitivity curves for various detectors,
taken from \cite{NoiseCurvesT1500293}.
}\label{fig:freq_evol}
\end{figure}

Figure~\ref{fig:waveforms} shows the plus polarization of the 
GW from three data sets. The purple line is the result 
extracted from the numerical simulation. Three characteristic phases 
are clearly identifiable. During the inspiral, the amplitude and frequency 
gradually increase until the maximal amplitude of the complex strain, $h = h_+ 
- i\, h_\times$, is 
reached at $t = t_\mathrm{merger}$. The following post-merger oscillation is 
characterized by an overall slowly decaying 
amplitude.
Figure~\ref{fig:freq_evol} shows the instantaneous frequency, 
$F = (2\pi)^{-1} d\phi / dt$, where $\phi$ is the GW
phase extracted as the argument of the complex strain $h$. 
It exhibits a characteristic modulation, with an initially large but rapidly 
damped amplitude. As we will show in Sec.~\ref{sec:remnant}, this modulation 
is an imprint of the remnant's radial oscillations. Such an imprint might be 
exploited in observations with next-generation 
instruments. Apart from the modulation, the frequency also shows a slow
drift towards higher frequencies. This correlates with a change in remnant 
compactness that will be investigated in Sec.~\ref{sec:remnant}. 
Once the BH is formed, 
the signal amplitude decays quickly while the frequency reaches the value that 
is 
expected for a BH with the mass and spin found in our simulations (this can only be 
observed briefly as the amplitude quickly becomes too small for numerical study).

The other two curves shown in Fig.~\ref{fig:waveforms} are predictions from 
waveform models commonly used in the LIGO and Virgo data analysis. In order to 
visually 
compare them to the numerical simulations and hybridize the waveforms, we 
aligned each model with the respective signal from our numerical 
simulation using the following procedure. First, after generating the model 
waveforms using the same masses as our numerical simulations and zero spins, we 
align the signals in time by minimizing the $\mathbb L_2$ norm of the difference 
between the phase velocities $\omega = 2 \pi F$, taken over the 
time interval where $\omega \in [3850, 6000] \usk\rad\per\second$. Second, we adjust 
the phase offset in the model such that the average phase difference between 
the numerical data and the models vanishes in the interval specified above. The 
only free choice in this procedure is the interval used for the alignment. 
It has to be chosen small enough to align the waveforms in the ``early'' 
inspiral of the numerical simulation. On the other hand, the size of the 
interval has to be large enough so that the frequency evolves significantly 
\cite{MacDonald:2011ne}. 
Otherwise the time shift would only be weakly constrained. The range defined 
above is an empirically found compromise that is shown as a shaded band in 
Fig.~\ref{fig:waveforms}.

One model used for comparison is the binary BH (BBH) model \texttt{IMRPhenomD} 
\cite{Husa:2015iqa, Khan:2015jqa} that does not incorporate tidal, finite-size 
effects of NSs. We would therefore not expect it to accurately 
describe the late inspiral and merger of a BNS.
However, the tidal effects for the cases at hand are small in the frequency
range used for the fitting, quite likely within the numerical error of the 
simulations. We stress that the aim of our study is not the accurate modeling
of the inspiral phase, which would require very high resolution (see, e.g.,  
\cite{Kiuchi:2020:084006}).

Just before merger, the 
BH model and NS simulation start to diverge significantly. 
A BBH with the same masses, following the alignment of the model and 
simulations used here, would perform about 1.5 orbits more than the BNS before 
merging. The most striking difference then, of course, is 
that the BBH forms a remnant BH immediately at 
merger, whereas our BNS mergers result a short-lived HMNS which emits strong GW
until it collapses to a BH. During merger, the signal shows an
amplitude minimum accompanied by a phase jump that is characteristic to 
BNS mergers \cite{Kastaun:2017}, but we observe none of the secondary minima/phase 
jumps which can sometimes be present. We will revisit this point in \Sref{sec:struct3d}.

The mass of the final BH that would result from the 
analogous BBH case can be computed using the fit to nonprecessing NR simulations 
by Varma et al \cite{Varma:2018aht}. The fit predicts that BBH mergers 
in the mass ratio $0.9$ and $1$ case produce remnants with mass  
$\MFBBHMEQUAL\,M_\odot$ and dimensionless spins of \CHIFBBHMNONEQ and 
\CHIFBBHMEQUAL, respectively. Somewhat surprisingly, this agrees within a 
few percent with the parameters of the BH formed in the BNS case, shown in 
Table~\ref{tab:outcome}.

Waveform models that include tidal
deformations of the NSs and the resulting effect on the binary's orbit are more 
appropriate for the systems we simulate here. As an example of current 
state-of-the-art models, we employ the \texttt{IMRPhenomD\_NRTidalv2} model 
\cite{Dietrich:2019kaq}
that adds an NR-informed description of the tidal dephasing on top of the 
BH model \cite{Dietrich:2017:1706.02969:arxiv}. 
The model is also shown in Fig.~\ref{fig:waveforms}.
While visually there is no difference to the BH 
model over the fitting region, the effect of the tidal phase corrections 
becomes visible as a gradual dephasing in the earlier inspiral. The impression 
that the dephasing between BH and tidal model seems to increase as one moves to 
earlier times is an artifact of aligning the models in the late inspiral. 
\texttt{IMRPhenomD\_NRTidalv2} does not attempt to model the merger and 
postmerger accurately; it simply decays rapidly beyond the contact frequency of 
the two NSs.

We use \texttt{IMRPhenomD\_NRTidalv2} to construct hybrid waveforms that 
cover the GW signal from the very early inspiral starting at $20\,\textrm{Hz}$ 
to the end of what was simulated numerically. We smoothly blend the inspiral 
model and the NR data over the same region that we used for aligning the signals 
in Fig.~\ref{fig:waveforms}. The boundaries of this interval $[t_1, t_2]$ 
inform a Planck taper function \cite{McKechan:2010:27h4020M},
\begin{align}
 \mathcal T(t) = \left \{ 
 \begin{array}{ll}
    0, & t \leq t_1 \\
    \left[ 1 + \exp\left( \frac{t_2 - t_1}{t - t_1} +  \frac{t_2 - t_1}{t - 
t_2}\right) \right]^{-1},  & t_1 < t < t_2 \\
1, & t \geq t_2
 \end{array}\right. ,
\end{align}
which we use to construct a $C^\infty$ transition with compact support of the 
form
\begin{align}
 X_\mathrm{hyb}(t) = \mathcal T(t) \, X_\mathrm{NR}(t) +  \left[ 1 - 
\mathcal T(t) \right] X_\mathrm{insp}(t) .
\end{align}Here, $X(t)$ stands for the amplitude or phase of the complex strain, which are 
hybridized individually. 

The hybrid waveforms are released as data files \cite{zenodo:4570825} with this article to facilitate 
exploratory data analysis studies. However, we caution that the accuracy of 
both the inspiral and NR data may not be sufficient for high-accuracy 
applications. Nevertheless, they may be used to estimate the order of 
magnitude at which differences in waveforms become measurable. As an example, 
we calculate the mismatch between the hybrids and the analytical waveforms 
shown in Fig.~\ref{fig:waveforms}. 

The mismatch quantifies the disagreement 
between two signals akin to an angle between vectors. We employ the standard 
definition of the mismatch,
\begin{align}
 \mathcal M (h_1 , h_2) &= 1 - \max_{\delta \phi, \delta t} \frac{\left \langle 
h_1  \vert h_2 \right \rangle}{\| h_1 \| \, \| h_2 \|}, \\
\left \langle 
h_1  \vert h_2 \right \rangle &= 4 
\operatorname{Re} \int_{f_1}^{f_2} \frac{\tilde h_1 (f) \, \tilde 
h_2^\ast(f)}{S_n(f)} df,
\end{align}
where $\tilde h(f)$ is the Fourier transform of $h(t)$, $^\ast$ denotes complex 
conjugation, $S_n(f)$ is the power spectral density of the assumed instrument 
noise, and the mismatch is minimized over relative time ($\delta t$) and phase 
($\delta \phi$) shifts between the two signals. $\|h \|^2 = \langle h \vert h 
\rangle$ is the norm induced by the inner product. As examples, we 
calculate mismatches using the noise curves provided in \cite{NoiseCurvesT1500293} for 
Advanced LIGO \cite{TheLIGOScientific:2014jea}, LIGO Voyager 
\cite{Evans:2016mbw} and the Einstein Telescope \cite{Punturo:2010zza}. For simplicity,  
we use the starting frequency of the hybrid, $f_1 = 
20\,\mathrm{Hz}$ in our calculations. Note that the 
assumed instruments are sensitive to lower frequencies, but as we mainly want 
to illustrate the effect of the merger and post-merger, our results are 
meaningful even for this artificially chosen starting frequency.

As we can see from the 
results in Table~\ref{tab:mismatches}, the hybrids disagree significantly more 
with the BBH model than with the tidal NS model. $\mathcal M_{\mathrm{BBH}}$ is 
dominated by the tidal effects of the inspiral, i.e., it reflects the 
difference between the tidal inspiral model chosen for hybridization and the 
BBH model. Note, however, that the mismatch is larger than it would be in a 
real parameter estimation study, where the masses and spins are not fixed, 
such that the BBH model could partly mimic tidal effects at the expense of 
biasing these parameters. On the other hand, by comparing the hybrids with the same 
inspiral waveform used for hybridization, we can quantify the effect of the 
merger and post-merger that is only present in the hybrid. Those mismatches for 
all assumed instruments are $\mathcal O(10^{-4})$. This might seem surprising at 
first, given that e.g., the Einstein Telescope is more sensitive than 
aLIGO. However, because the mismatch is based on the normalized inner product, 
its value is determined by the relative weight between low and high frequencies 
as given by the noise spectral density, and not by the instrument's overall 
sensitivity. Choosing the same lower cutoff frequency $f_1 = 
20\,\mathrm{Hz}$ for all instruments exaggerates this effect.

\begin{table}
\caption{Comparison of our hybrid waveforms with either the the
    BBH model (\texttt{{IMRPhenomD}}) (first row) or the tidal inspiral model (\texttt{{IMRPhenomD\_NRTidalv2}}) (remaining rows). 
    We present mismatches $\mathcal M$ assuming instrument noise curves for aLIGO second observing run O2, aLIGO design 
    sensitivity, LIGO Voyager and the Einstein Telescope. 
    The last two rows indicate at what SNR the hybrid and the tidal inspiral model would 
    distinguishable at the 90\% credible level (see text), and at which distance this SNR would be achieved for optimally oriented binaries.}
\begin{ruledtabular}
\begin{tabular}{lrrrrrrrr}
&
\multicolumn{2}{c}{O2}&
\multicolumn{2}{c}{aLIGO}&
\multicolumn{2}{c}{Voyager}&
\multicolumn{2}{c}{ET} \\&
{\small \texttt{Q09}}&
{\small \texttt{Q10}}&
{\small \texttt{Q09}}&
{\small \texttt{Q10}}&
{\small \texttt{Q09}}&
{\small \texttt{Q10}}&
{\small \texttt{Q09}}&
{\small \texttt{Q10}}
\\\hline 
$\mathcal{M}_{\mathrm{BBH}} \; [10^{-4}] $&
$51$&
$52$&
$120$&
$110$&
$49$&
$49$&
$96$&
$98$
\\
$\mathcal{M}_{\mathrm{BNS}} \; [10^{-4}]$&
$0.8$&
$1.7$&
$2.8$&
$4.6$&
$0.7$&
$1.6$&
$1.7$&
$3.3$
\\
$\mathrm{SNR}_{90}$&
$130$&
$91$&
$69$&
$54$&
$140$&
$93$&
$89$&
$64$
\\
$D_{90} \; [\mathrm{Mpc}]$&
$13$&
$19$&
$48$&
$62$&
$100$&
$160$&
$380$&
$530$
\end{tabular}
\end{ruledtabular}
\label{tab:mismatches}
\end{table}
 
We can appropriately account for the actual detector sensitivity by relating 
\emph{measurability} of a difference between two signals with the 
\emph{signal-to-noise ratio} (SNR). Following the derivation in 
\cite{Baird:2012cu, Ohme:2013nsa}, one finds that the difference between 
the hybrid and the inspiral model is indistinguishable at the $p$-probability 
level if
\begin{align}
 \| h_{\mathrm{hyb}} - h_{\mathrm{BNS}}\|^2 < \chi^2_k (p), \label{eq:wf_diff}
\end{align}
where $\chi^2_k (p)$ is a number derived from the $\chi^2$-distribution 
with $k$ degrees of freedom at which the cumulative probability is $p$. Here 
we consider the question at what SNR the waveform differences are 
distinguishable at a 90\% level for a one-dimensional distribution. 
Using the corresponding values $p=0.9, k=1$ results in $\chi^2_k(p) = 2.71$. Expanding the 
left-hand side of Eq.~(\ref{eq:wf_diff}) for small $\mathcal M$, 
we finally obtain
\begin{align}
 \min_{\| h_{\mathrm{BNS}} \|} \| h_{\mathrm{hyb}} - h_{\mathrm{BNS}}\|^2 
\approx 2 \| h_{\mathrm{hyb}}\|^2 \mathcal M < \chi^2_k (p),
\end{align}
which allows us to estimate the critical SNR ($\| h_{\mathrm{hyb}}\|$) required 
to distinguish the two signals given their mismatch. The result is shown in the 
third row of Table~\ref{tab:mismatches}. Further assuming an optimally oriented 
source overhead the detector, we can calculate the luminosity distance at which 
the critical SNR is achieved. This last row in Table~\ref{tab:mismatches} 
follows the expected trend: more sensitive, future-generation instruments would 
be able to measure the difference between our hybrid and the inspiral tidal 
model out to a greater distance. We note that the SNR contained in the 
hybrid waveform beyond the merger frequency accounts for most of the mismatch 
we calculate, i.e.,
\begin{align}
 \frac{\| h_{\mathrm{hyb}}(f > f_\mathrm{merger}) \|^2}{\| h_{\mathrm{hyb}} 
\|^2} \sim 2 \mathcal M \sim \mathcal O \left ( 10^{-4} \right).
\end{align}
While the mismatch and overall SNR may be affected by our choice of 
lower cutoff frequency $f_1$, the distance we quote is dominated by the 
post-merger and largely independent of the specific choice of $f_1$. Hence, it 
defines the volume in which the 
specific post-merger signal from our simulations is distinguishable from the
tidal model without post-merger contribution. A study of similar 
questions 
was published in \cite{Dudi:2018jzn}. We stress that we do not
address the more complicated question at which distance the presence of an unknown 
post-merger signal can be observed.

As a final application of our waveform comparison, we use the inspiral data to 
estimate the eccentricity of our NR simulations. We do this by comparing the 
frequency evolution $\omega(t) = d\phi/dt$ of the NR data with the 
quasi-circular data of the \texttt{IMRPhenomD\_NRTidalv2} model. The residual 
difference
\begin{align}
 e_\omega = \frac{\omega_\mathrm{NR} - \omega_\mathrm{circ}}{2 
\omega_\mathrm{circ}}
\end{align}
can be fit by a sinusoidal oscillation added to a small linear drift that 
absorbs any inaccuracies in the alignment. The amplitude of the oscillatory 
part of 
$e_\omega$ characterizes the eccentricity of the NR simulation 
\cite{Mroue:2010re}. We find initial eccentricities  \ECCMEQUAL and 
\ECCMNONEQ for the equal-mass and mass ratio $0.9$ simulation, respectively.

\subsection{Radial Remnant Profiles}
\label{sec:remnant}

\begin{figure}
\includegraphics[width=0.95\columnwidth]{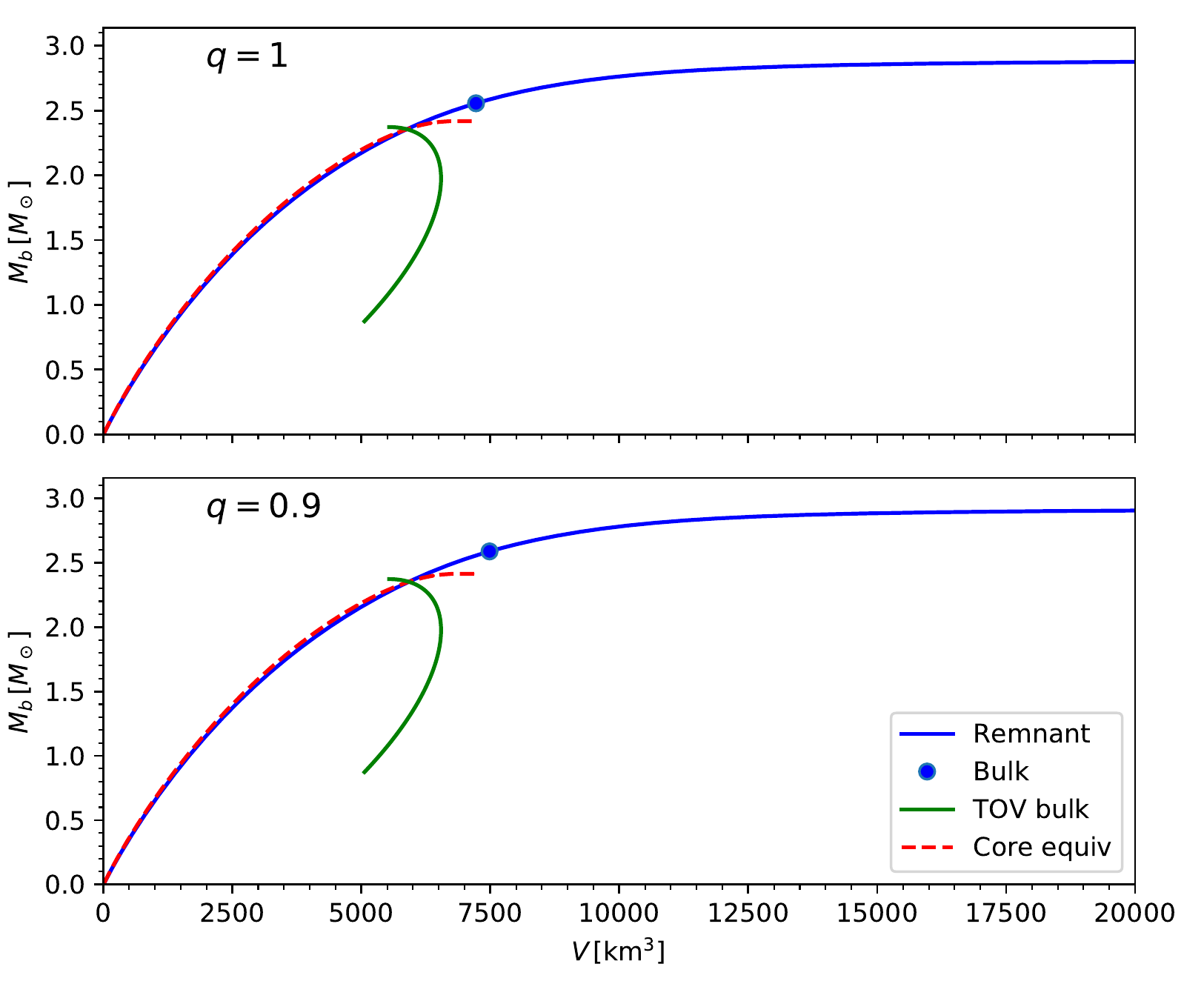} 
\caption{Total baryonic mass contained inside surfaces of constant 
density (blue curve) versus proper volume contained within the same 
surfaces. The top panel shows the remnant for the equal mass case,
the bottom panel for the unequal mass system, both at a time 
$1\usk\milli\second$ before apparent horizon formation.  
The dot marks bulk mass and bulk volume of the remnant (see text).
For comparison, we show bulk mass versus bulk volume (green line) for 
the sequence of nonrotating NS following the same EOS as the BNS initial 
data, starting at mass $0.9 \usk M_\odot$ up to the maximum bulk mass.
The intersection with the remnant mass-volume curve defines the 
core equivalent TOV model, for which we show the mass-volume
relation as well (dashed red curve).
}
\label{fig:mass_volume}
\end{figure}

We begin our discussion of the remnant structure with the average 
radial mass distribution shortly before the onset of collapse.
For this we use the measure introduced in \Sref{sec:diag}. 
\Fref{fig:mass_volume} shows the profile of baryonic mass contained
within isodensity surfaces versus the proper volume within the same 
surfaces. We also mark the bulk region defined in \Sref{sec:diag}.
As shown in \Fref{fig:ns_disk_xz}, there is a smooth transition 
between remnant core and surrounding disk. This is also reflected
in the mass-volume profile.

It is instructive to compare this profile to those obtained for
nonrotating NS with same EOS as the initial data.
In a previous work \cite{Kastaun:2016}, we introduced a method to find a
nonrotating NS model (with same EOS as the initial data) 
for which the profile of the core resembles 
the one of the remnant. To this end, we compute the bulk mass versus
bulk volume for the whole sequence of nonrotating NS and find the
intersection with the remnant mass-volume profile, provided that 
it does intersect. 

The sequence is shown in \Fref{fig:mass_volume} and just barely
intersects the remnant profile, near the maximum mass NS model. 
The figure also shows the 
mass-volume profile of the corresponding NS model, which we refer to 
as core-equivalent TOV model. It agrees remarkably well with the
merger remnant profile within the whole bulk of the NS. Close
to the NS surface, the two profiles naturally start to deviate, with 
the merger remnant profile smoothly extending to the debris disk.

\begin{figure}
\includegraphics[width=0.95\columnwidth]{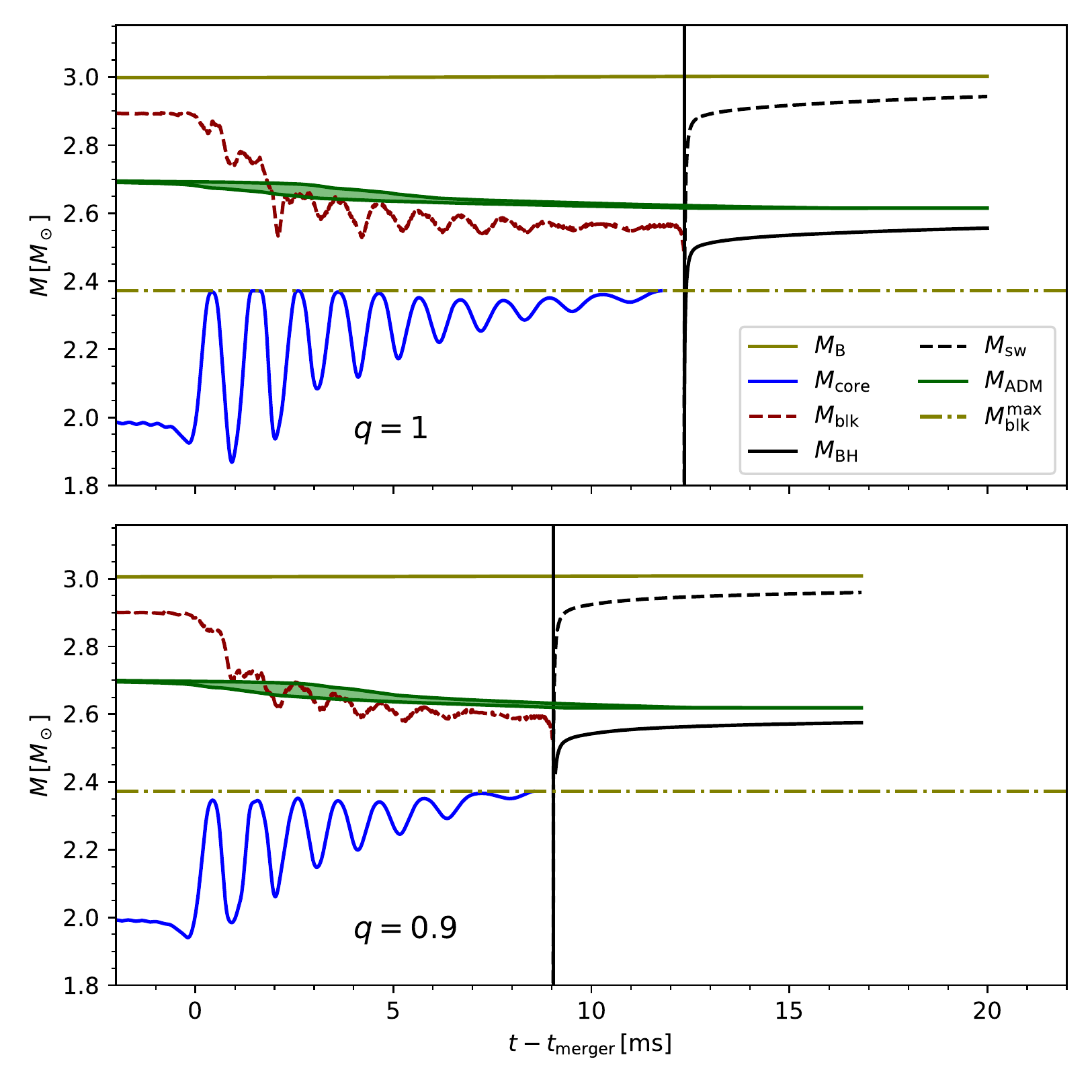} 
\caption{Evolution of masses for the equal mass case (top panel) and the unequal
mass case (bottom panel). The dashed red curve shows the bulk baryonic mass,
the solid blue curve the bulk baryonic mass of the core TOV equivalent (see main text), 
and the horizontal dash-dotted line the maximum bulk baryonic mass of nonrotating 
NS following the same EOS as the BNS initial data (SFHO). The solid
black curve shows the gravitational mass of the BH (estimated from the apparent 
horizon) and the vertical line the time of first apparent horizon detection.
The solid olive curve shows the total baryonic mass present in the computational domain
up to BH formation, and a constant afterwards. The
dashed black curve shows the baryonic mass swallowed by the BH.
The green curve shows the ADM mass, where the shaded area is the contribution 
attributed to the energy of GW radiation within the computational domain.
}\label{fig:eqtov_evol}
\end{figure}

The time evolution of the core equivalent mass is shown in \Fref{fig:eqtov_evol}, whereas 
the evolution of the bulk density is depicted in \Fref{fig:dens_evol}.
We find large initial oscillations which are almost completely damped until collapse.
Simultaneously with the damping, we also observe a drift towards larger equivalent core mass
and larger bulk density.

At some point, the remnant bulk density exceeds the maximum bulk density of 
TOV solutions and also no core equivalent NS can be found anymore. 
Collapse sets in within less than  $1 \usk\milli\second$ after this point.
This behavior agrees well with earlier results \cite{Ciolfi:2017:063016,Endrizzi:2018} 
obtained for different systems. 
The mounting number of simulation results without any counterexample
adds weight to the conjecture that a HMNS collapses as soon as it does not allow 
for a core equivalent TOV model anymore. 

It should however be mentioned that the above conjecture is disregarding 
brief violations due to oscillations.
For the $q=1$ case, the core is slightly too compact to allow a stable TOV core 
equivalent for a very brief time already during the first oscillations after 
merger (this is hardly visible in the figure). 

Currently, it is up to 
speculation if one should expect collapse when this limit is briefly 
exceeded dynamically. Our original conjecture for the
collapse criterion is motivated by quasi-stationary systems. It is however 
worth noting an example where the limit was almost reached during the initial 
oscillations without any collapse, presented in \cite{Ciolfi:2017:063016}. 
The mass of this model was known to be just below the estimated threshold for prompt 
collapse for the given EOS (APR4). It seems likely 
that the merger studied here is also very close to prompt collapse.

\begin{figure}
\includegraphics[width=0.95\columnwidth]{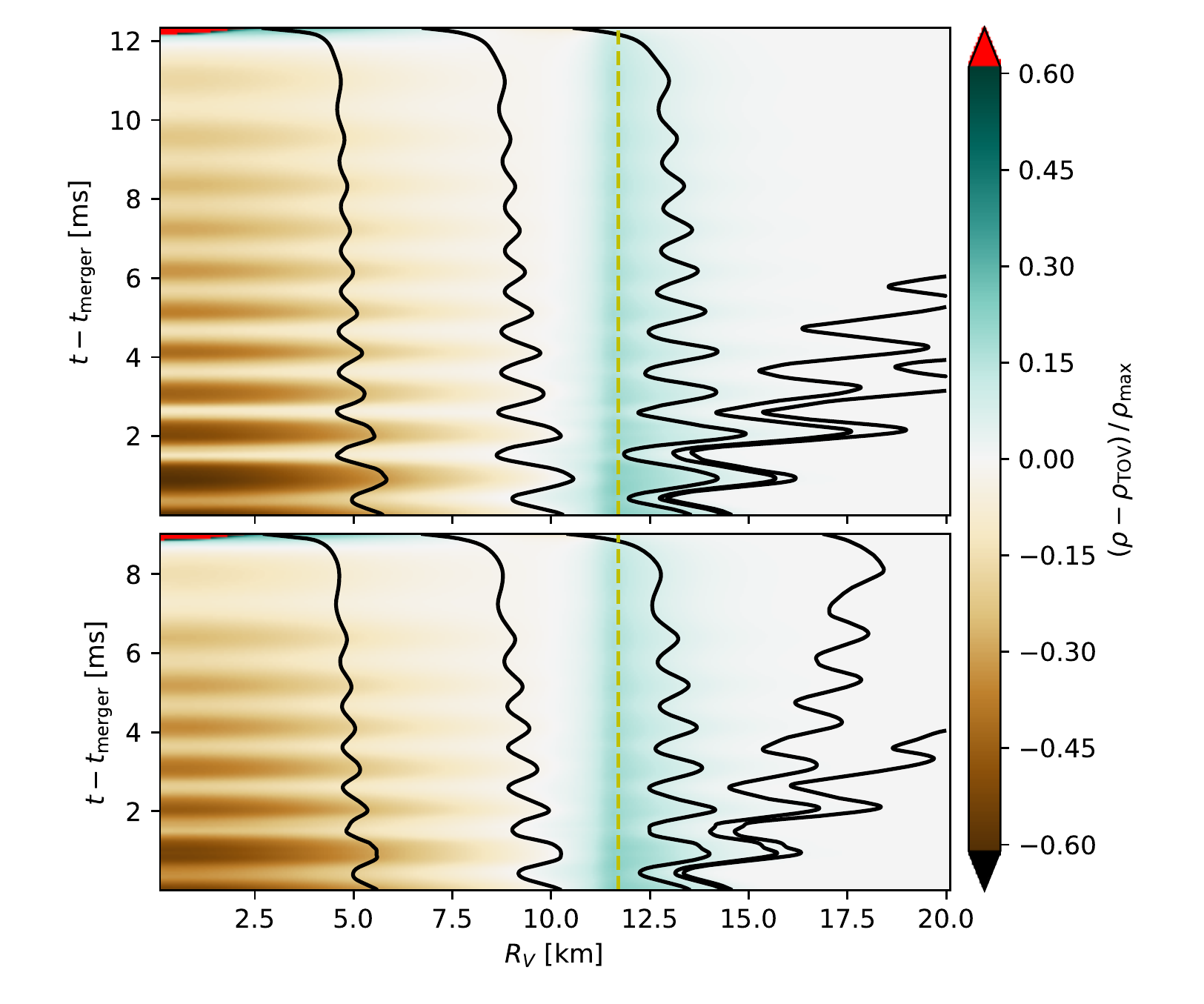} 
\caption{Evolution of radial mass distribution in the merger remnant
for the equal mass case (top panel) and unequal mass case (bottom panel).
The solid black curves show the time evolution of the proper volume 
within isodensity surfaces containing fractions 
$0.1,0.5,0.9,0.97$, and $0.98$ of the total baryonic mass.
The volume is given in terms of volumetric radius $R_V$, the radius
of an Euclidean sphere of equal volume.
We further compare the remnant to the spherical NS solution of maximum 
bulk mass. For both, we compute the density as function
of proper volume within isodensity surfaces. The color scale
represents the difference of density at given volumetric radius, 
normalized to the maximum density of the spherical NS. The volumetric 
surface radius of the latter is shown as dashed vertical line. 
}\label{fig:radial_evol}
\end{figure}

Next, we turn to study the time evolution of the radial mass distribution.
Since the average radial distribution in the late core is very close to the one of 
the maximum mass spherical NS, it is natural to subtract the latter.
\Fref{fig:radial_evol} shows the resulting differences in density profile in 
a time-radius diagram.
As one can see, the deviations from the maximum mass TOV profile are 
larger at first, up to 60\% of the maximum density, 
and also show large oscillations. A noteworthy property
of the oscillations is that the density in the core does not significantly 
exceed the TOV model until shortly before collapse.

To further investigate the oscillations, we also show the volumetric radius 
of surfaces containing fixed amounts of baryonic mass in \Fref{fig:radial_evol}.
The oscillation of the occupied proper volume provides a definition for an average 
radial displacement associated to those oscillations. Again, we see a strong damping of the 
oscillations. 

In order to compute the frequency of the radial oscillation, we determine the
maxima and minima of the bulk density shown in \Fref{fig:dens_evol}, after subtracting a quadratic
fit to remove the drift. We then compute frequencies from the time
between adjacent pairs of maxima as well as pairs of minima. The result is 
shown in \Fref{fig:freq_evol}. 
We find that the radial oscillation frequency decreases when approaching 
collapse. 

We recall that for a nonrotating NS, collapse occurs when the square 
of the frequency of the radial quasinormal mode crosses zero. 
Under the assumption that the collapse 
mechanism for the merger remnant is the same, one would expect the radial oscillation
frequency to approach zero as well. 
The evolution in \Fref{fig:freq_evol} is compatible with this picture, 
although we cannot determine the oscillation frequency arbitrary close to the 
collapse (since we measure the oscillation frequency by distance between extrema).

Angular momentum conservation suggests that the radial oscillation should cause
a modulation of the overall rotation, and therefore of the GW frequency.
Indeed, the GW frequency shown in \Fref{fig:freq_evol} shows minima and maxima
that correlate with those of the bulk density. The modulation frequency obtained 
from minima and maxima of the GW frequency agrees very well with the radial
oscillation frequency obtained from the density, and the modulation strength
decreases with the radial oscillation amplitude.

As shown in \Fref{fig:radial_evol}, the volume occupied by isodensity surfaces
containing fixed baryonic masses shows a slow decrease in the core. In this sense, 
the core is shrinking.
Besides the core, \Fref{fig:radial_evol} also shows the transition zone
between remnant and disk. 
Here, the figure clearly shows an expansion of the isodensity surfaces of fixed mass.
Even though 
the isodensity surfaces for the disk are not spherical anymore, our measure
tells us that they occupy more space. This rules out mass accretion onto the remnant 
as a cause for the increasing compactness of the core.

\subsection{Three-dimensional Remnant Structure}
\label{sec:struct3d}

After studying the average radial mass distribution, we now turn to investigate
the overall structure of the 3D fluid flow inside the remnant. 
Conceptually, we decompose the dynamics into a rotation with slowly drifting 
angular velocity, a quasi-stationary flow pattern, and subdominant contributions
such as quasi-radial oscillations. 

To extract the quasi-stationary part of the flow in the rotating frame, we 
employ a complex postprocessing chain as follows.
First, we select a time window for averaging. For each of the 3D datasets 
saved during the simulation at regular intervals within the window, we
first load 3D mesh-refined simulation data. 
This data is resampled onto a regular grid uniform in simulation coordinates 
that is covering the region of interest.
Next, we apply the coordinate transformation into the corotating postprocessing 
coordinates described in \Sref{sec:gauge}. During this step, we resample again
onto a regular grid, this time uniform in the postprocessing coordinates.
We also compute the Jacobian of the transformation in order to transform vectors 
correctly.
To account for the time-dependent transformation of spatial coordinates, we further 
compute a new shift vector with the corresponding corrections.
In this fashion, we compute the quasi-stationary density $\bar{\rho}$ and fluid velocity 
with respect to the corotating postprocessing coordinates, $\bar{w}^i$.

For visualization purposes, we compute the integral curves of the coordinate 
velocity $\bar{w}$. If the flow pattern were truly stationary, those curves would correspond
to fluid trajectories. Because the structure is slowly changing and because we average
out oscillations, the integral curves do not agree exactly with trajectories. That said,
they do represent a good measure for the overall structure of the fluid movement.

\begin{figure*}
\includegraphics[width=0.99\textwidth]{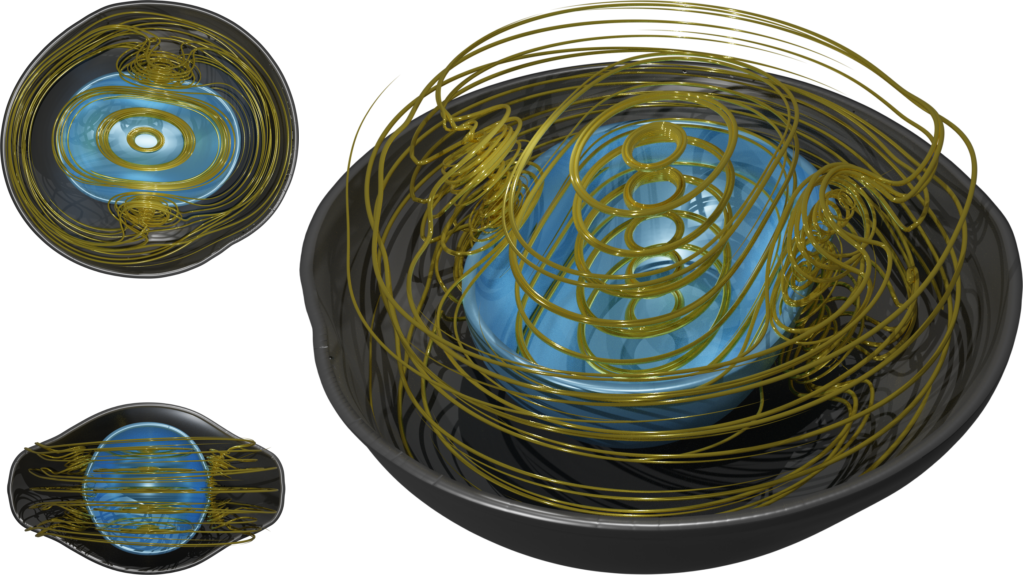}
\caption{Quasi-stationary part of the remnant structure, in the frame corotating with $m=2$ 
density perturbation of remnant. 
The visualization represents the average over a time window 
$7\pm1 \usk\milli\second$ after merger ($2\usk\milli\second$ before collapse)
for the $q=0.9$ case.
The inner and outer surfaces (cut open) mark mass densities $0.3$ and $0.01$ 
of the maximum density, respectively.  
The wires represent integral lines of the averaged velocity field, shown inside 
the dense region enclosed within the outer surface.
The top left rendering shows the remnant from a perspective along the rotation axis,
the bottom left one from the side, along the longer axes of the deformed core, looking
onto a meridional plane that cuts through the secondary vortices. 
}\label{fig:remnant3d}
\end{figure*}

\Fref{fig:remnant3d} shows the integral curves together with two isodensity surfaces
of $\bar{\rho}$ around $7\usk\milli\second$ after merger for the 
unequal mass case.
A prominent feature visible in the figure is the presence of two secondary vortices.
Such vortices seem to be a generic feature, which we have observed in previous
works \cite{Kastaun:2016,Kastaun:2017,Ciolfi:2017:063016} that studied the fluid 
flow within equatorial plane. The 3D results shown in \Fref{fig:remnant3d} demonstrate 
how those vortices extend above and below the equatorial plane. 
We find that the direction of the fluid flow has negligible vertical 
components, except for the region within the secondary vortices. Even there,
the absolute velocities are small. This suggests that mixing of matter in the 
vertical direction can probably by neglected.

The figure also shows that the inner fluid flow is still strongly non-axisymmetric.
We recall that our coordinates are constructed such that a physically axisymmetric 
system would also appear axisymmetric in the postprocessing coordinates.
The deformation of the fluid flow correlates with a strong elliptical deformation of the 
inner isodensity surface shown in the figure. 
The isodensity surface outside the secondary vortices
is deformed less strongly, but exhibits some bumps which seem to be related to the 
secondary vortices. 

Notably, the bumps in the outer regions are oriented nearly orthogonal to the 
deformation of the core. This is relevant for the GW signal, since the corresponding 
quadrupole moments in the rotating frame have different sign. The resulting GW signal 
is then the difference of two contributions. This might explain why numerical simulations 
sometimes exhibit pronounced secondary minima in the post-merger signal that are accompanied by
a phase jump, as discussed in \cite{Kastaun:2017}. The relative amplitudes of the two contributions
can change, which might result in a zero-crossing of the quadrupole moment in the rotating 
frame. We reserve the quantitative discussion of this effect for future work but point out
that, even though the deformation of the outer regions seems less pronounced and is 
located in less dense regions, this might be compensated by the quadratic radial factor in
the qudrupole moment and the cubic radial factor from the volumes involved.

\begin{figure*}
\includegraphics[width=0.98\textwidth]{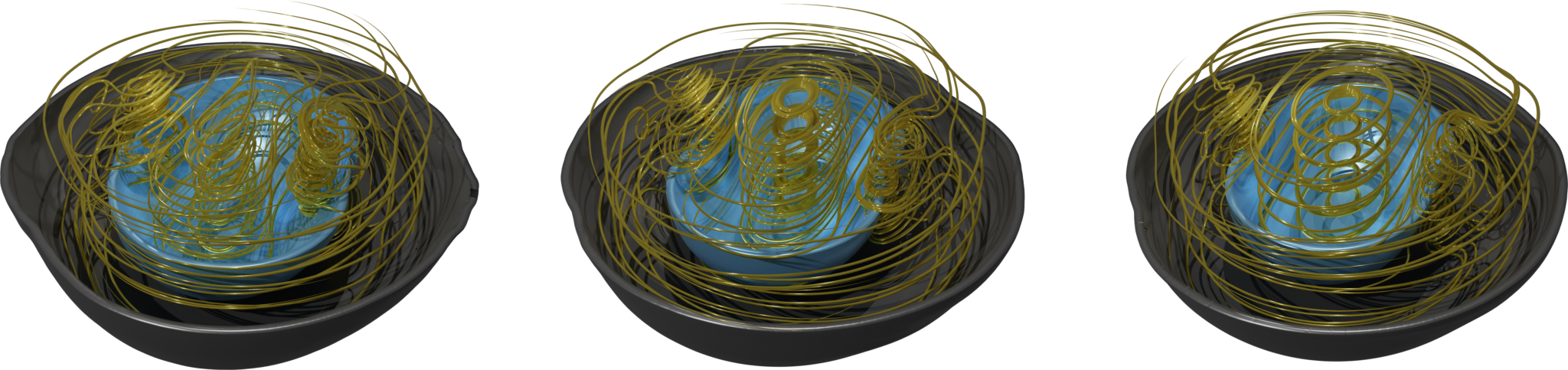} 
\caption{Remnant structure visualized as in \Fref{fig:remnant3d}, but for 
three different time windows.
From left to right:
$3\pm1 \usk\milli\second$ after merger, 
$5\pm1 \usk\milli\second$ after merger, and
$7\pm1 \usk\milli\second$ after merger ($2\usk\milli\second$ before collapse).
The camera distance remains constant to allow size comparison. 
}\label{fig:flow3devol}
\end{figure*}

In \Fref{fig:flow3devol}, we compare the remnant structure at time
shown in \Fref{fig:remnant3d} to times shortly after merger and shortly
before collapse. Although there are some differences, the non-axisymmetric 
deformation and the secondary vortices stay prominent right until 
collapse. This corresponds to a large GW amplitude sustained until 
collapse, which was shown in \Sref{sec:gw}.

\subsection{Differential Rotation}
\label{sec:diffrot}

\begin{figure}
\includegraphics[width=0.98\columnwidth]{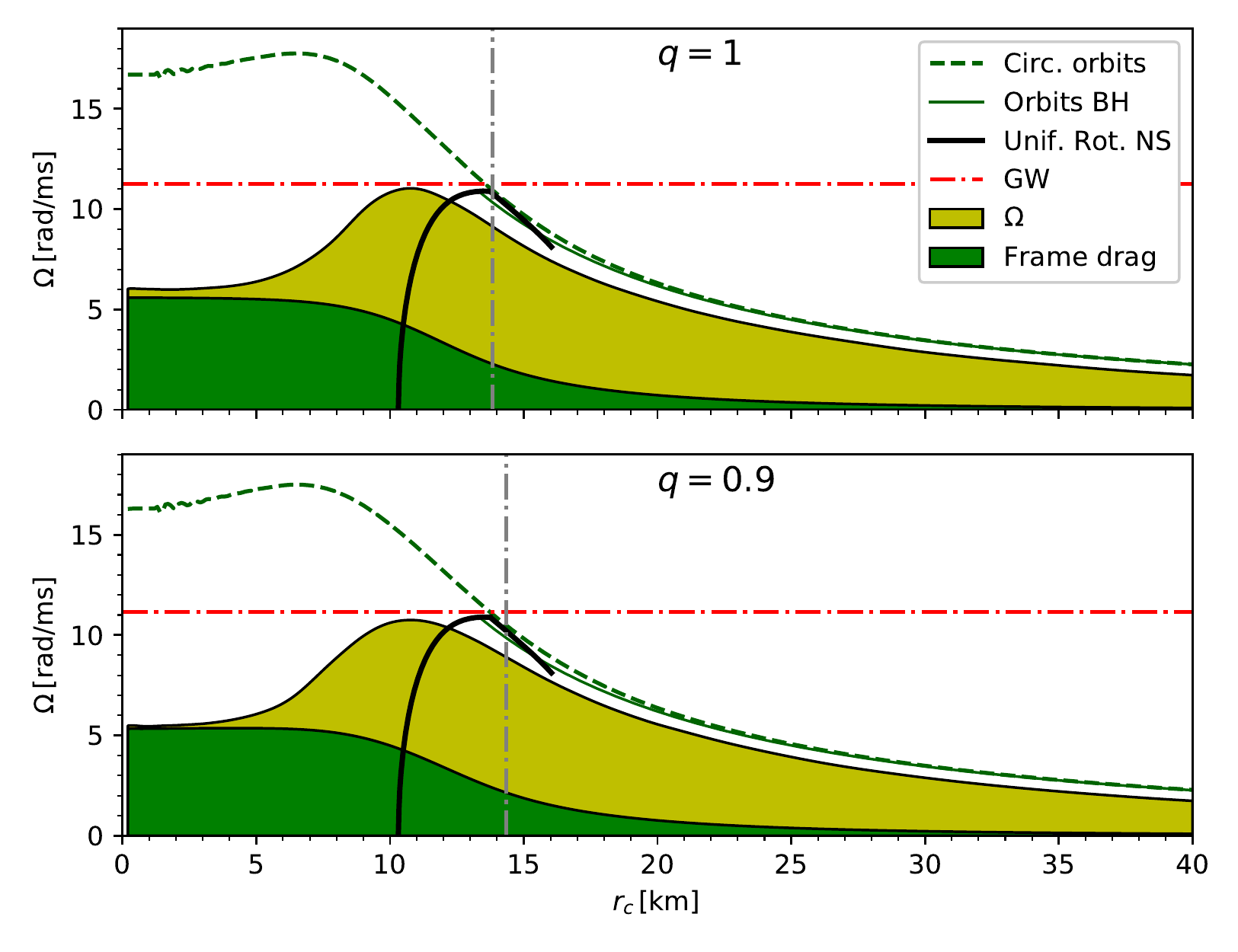} 
\caption{Rotation profile of the merger remnant $1\usk\milli\second$ before  
collapse, for mass ratios $q=1$ and $q=0.9$. The filled curve shows the 
angular velocity (as seen from infinity) and the frame dragging contribution, 
in the equatorial plane  versus circumferential radius. The data has been
averaged in $\phi$-direction and over a time window $\pm 0.5 \usk\milli\second$. 
For comparison, we show the angular velocity
of test particles in prograde circular orbit (dashed curve),
and the same for particles orbiting the spinning BH  
formed after collapse (solid green curve). The solid black curves
show the rotation rate and surface radius of uniformly rotating
supramassive NS, either at mass shedding limit (right curve) or at minimal
angular momentum (left curve).  
The vertical line marks the radius where the $\phi$-averaged mass density
in the equatorial plane falls below 5\% of the central value.
The horizontal line marks half of the GW angular frequency.
}\label{fig:rot_prof}
\end{figure}

We now discuss the rotation profile of the remnant. Although the fluid flow
is decidedly non-axisymmetic, it is instructive to study the axisymmetric 
part obtained by averaging in azimuthal direction. 
We start with the rotation profile in the equatorial plane. The 
azimuthal average  with respect to the post-processing coordinates is 
depicted in \Fref{fig:rot_prof}. The profile shows the same generic 
behavior found for many different systems \cite{Ciolfi:2017:063016,
Kastaun:2015:064027,Kastaun:2016,Kastaun:2017,Ciolfi:2019:023005,Endrizzi:2018,
Endrizzi:2016:164001,Hanauske:2016}. The rotation of the core is  
comparatively slow with respect to infinity, and it is rotating very 
slowly with respect to the local inertial frame. Further out, the rotation 
rate exhibits a maximum.

As in previous works, we find that this maximum average rotation rate 
agrees well with half of the instantaneous GW frequency.
This agreement has already been observed before \cite{Kastaun:2015:064027,
Kastaun:2016,Kastaun:2017, Endrizzi:2016:164001, Ciolfi:2017:063016, Ciolfi:2019:023005} 
and we are not aware of a counterexample. Although it is too early to generalize, the
indications accumulate that this relation might be typical.

The evolution of maximum rotation rate and half the GW frequency is shown in 
\Fref{fig:freq_evol}.
The frequencies do not just coincide at the time shown in \Fref{fig:rot_prof}.
Clearly, they agree well throughout most of the post-merger phase. 
However, a few $\usk\milli\second$ after the time $t_\mathrm{merger}$ of peak 
GW amplitude, the system is still in the process of merging.
The computed maximum rotation rate is not meaningful during this period 
because during this phase it corresponds to the shear component near the origin. 
Consequently the correlation to the GW frequency is not present.

At larger radii, the rotation rate slowly approaches the Kepler rate
as the remnant transitions into the disk. Since there still is a pressure
gradient in the disk, the rotation is slightly slower than the orbital
velocity.
For comparison, we also plot the orbital velocity profile for the BH
present shortly after collapse (mass and spin are given in \Tref{tab:outcome}). 
Naturally, it agrees well with the orbital velocity before collapse.
Surprisingly, the orbital frequency of the innermost stable circular
orbit agrees with the maximum rotation rate and half GW frequency before collapse.
We are not aware of any reason this should be the case, and it might well be
a numerical coincidence.

\Fref{fig:rot_prof} also shows the radius and rotation rate for two sequences
of uniformly rotating supramassive NS with same EOS as the initial data. One 
sequence is given by the models at mass shedding limit, and the other 
by models with the minimum angular momentum required to allow stable solutions.
We find that the radius of the maximum mass model is very close to the innermost 
stable circular orbit of the final BH. We also observe a close match between 
the rotation rate of the maximum mass model and the maximum of the remnant rotation 
rate profile. This might be a coincidence. Nevertheless, it should be noted
that such a relation would be extremely useful, since it would allow to predict the 
GW frequency of a HMNS directly before BH formation from the EOS alone, without even 
using the total mass of the system.

\begin{figure}
\includegraphics[width=0.98\columnwidth]{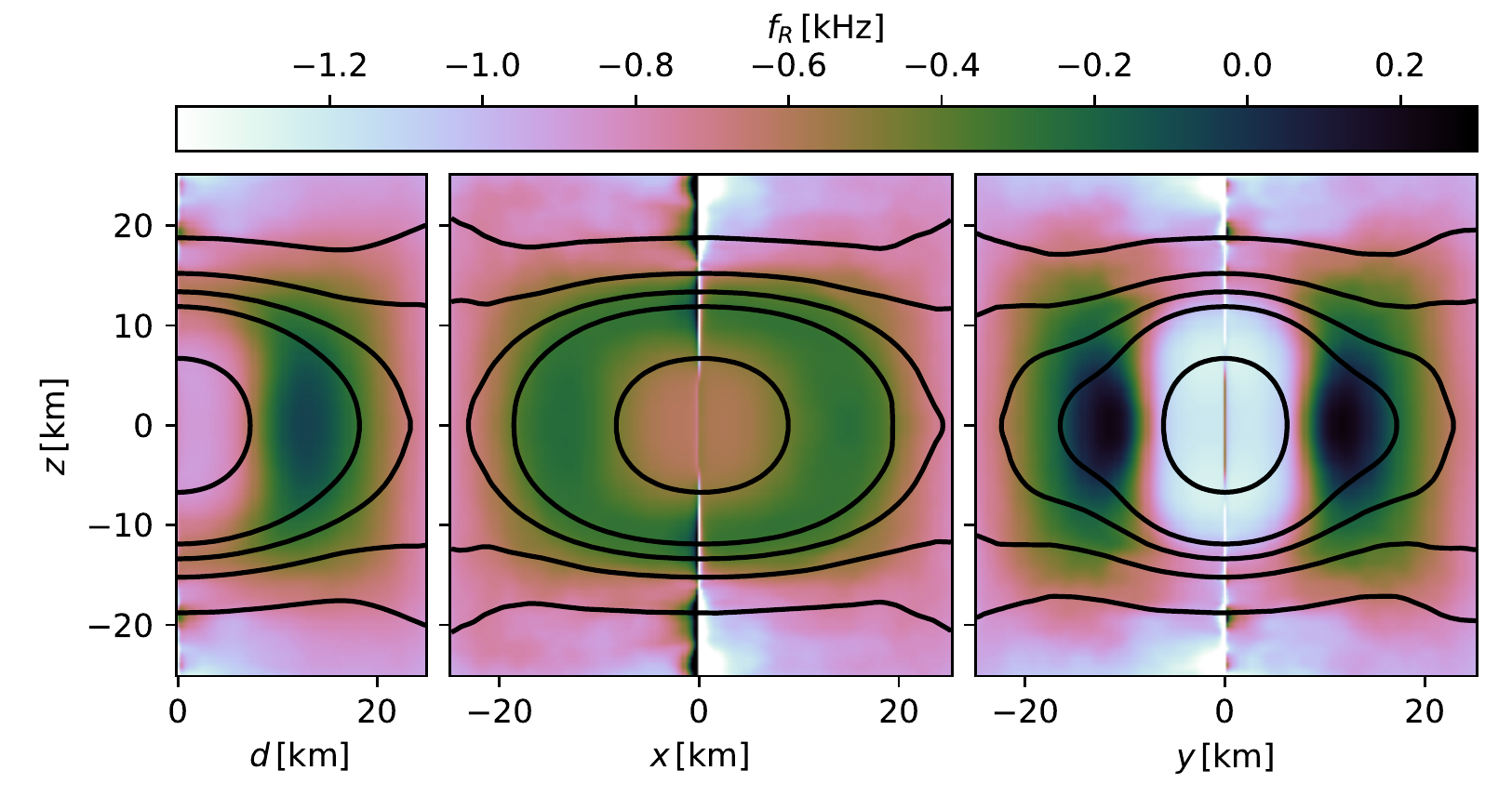} 
\caption{Twodimensional differential rotation profile of the remnant for mass ratio $q=0.9$,
averaged over the time window $7\pm1 \usk\milli\second$ after merger. 
The color scale denotes the rotational frequency in the 
coordinate system corotating with the perturbation pattern. 
Negative values signify that (in the inertial frame) the fluid is rotating more 
slowly than the pattern.
The left panel shows the average in $\phi$-direction versus cylindrical radius 
and z-coordinate. The right panel shows a cut in $(y,z)$-plane, 
the $y$-axis being roughly aligned with the secondary vortices shown 
in \Fref{fig:remnant3d}. The middle panel shows the $(x,z)$-plane instead.
The solid curves mark isodensity contours (in the left panel with respect 
to $\phi$-averaged density). Radial distances in the equatorial plane and 
distances along the z-axis are both proper distances, allowing
to asses the oblateness of the remnant.
}\label{fig:vphi_2d}
\end{figure}

We now turn to the rotation profile outside the equatorial plane.
The left panel of \Fref{fig:vphi_2d} shows the azimuthal average
of rotation rate as function of the distance $d$ to the $z$-axis and of 
the $z$-coordinate.
The rotation rate is computed with respect to a straight rotation axis 
orthogonal to the equatorial plane. 
The centers of rotation on each plane parallel to the equatorial plane 
nearly coincide with this line, but do not form a perfectly straight 
line. This misalignment is visible in the plot as artifacts close to the axis,
even though the underlying velocity field is smooth.

The rotation rate in the corotating postprocessing frame is mostly negative.
The region with zero rotation in this frame corresponds to the maximum rotation rate
in the inertial frame (compare \Fref{fig:rot_prof}).
Inside the remnant core, we find that the profile mainly depends on $d$ and 
less on $z$. Along the axis, we also observe some differential rotation 
in $z$-direction outside the dense regions.

From the 3D visualization \Fref{fig:remnant3d}, we already know that the fluid flow shows 
pronounced deviations from axisymmetry. This can also be seen in the middle and 
right panels of \Fref{fig:vphi_2d}. Those show the rotation rate in two 
meridional planes orthogonal to each other, one of which (right panel) is 
passing through the secondary vortices visible in \Fref{fig:remnant3d}.

The vortices themselves are stationary in the rotating postprocessing 
frame, and their own rotation is prograde with respect to the remnant.
The local rotation rate (w.r.t. inertial frame) inside the vertices exceeds the 
rotation rate of the dominant density perturbation on the vertex side 
opposite to the remnant center. The meridional plane crossing the 
vortices exhibits stronger gradients of rotation rate than the orthogonal 
meridional plane shown in the middle panel.

The local deviation of the flow from axisymmetry correlates with 
a nonaxisymmetric perturbation of the density. This can be seen
in the overlaid isodensity contours in \Fref{fig:vphi_2d}.
The right panel depicting the cut through the vortices shows a 
slightly prolate core. Further out, the isodensity contours are
not simple ellipsoids but exhibit an equatorial bump. In contrast,
the same density contour in the middle panel is nearly ellipsoidal 
and the core is slightly oblate.

\subsection{Disk Vorticity}
\label{sec:vorticity}

Besides the fluid flow inside the hypermassive NS, we are also interested in the
dynamics of the surrounding disk. As already shown in \Fref{fig:remnant_xyt}, the disk
is subject to continuous strong perturbations until the BH is formed. We have also
shown in \Fref{fig:radial_evol} that matter migrates from the NS into the disk.
One can therefore expect an impact on the fluid flow, causing deviations from 
a quasi-Keplerian disk.

The velocity field in the disk is dominated by the quasi-Keplerian velocity 
profile shown in \Fref{fig:rot_prof}. In order to get a more detailed picture, 
we compute the fluid vorticity in three dimensions, which, being a differential
expression, is more sensitive to local deviations.
We recall that the most appropriate vorticity measure in the relativistic case 
is given by $\partial_i (hv_j) - \partial_j(hv_i)$. However,
since metric and enthalpy are not saved as 3D data in our simulations, we instead 
compute the ordinary curl $\vec{\nabla} \times \vec{w}$, where 
$w^i = \alpha v^i - \beta^i$ is the fluid advection speed
with respect to the simulation coordinates, and $\vec{\nabla}$ refers to the 
ordinary partial derivatives.
We also do not use our usual postprocessing coordinates because they are not available 
after BH formation. This simple vorticity measure is sufficient 
for the following qualitative discussion of local shear, but not suitable
for a study of vorticity conservation.

\begin{figure*}
\includegraphics[width=0.98\textwidth]{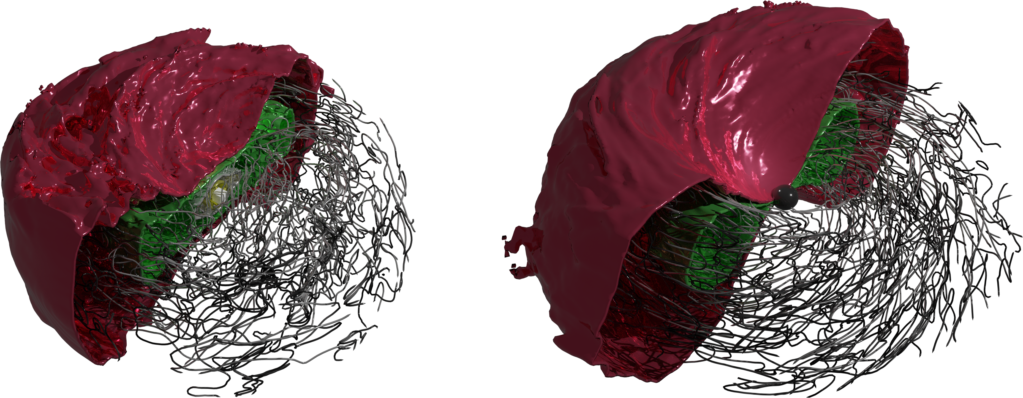} 
\caption{Vorticity lines inside the disk $2 \usk\milli\second$ before (left) 
and $5 \usk\milli\second$ after BH formation (right), for the $q=0.9$ case.
The black surface in the right panel marks the apparent horizon.
The cut-open solid surfaces in the left panel mark densities
$4.0\times 10^{10}, 2.2\times 10^{11}, 1.7\times 10^{13}$, and 
$5.0\times 10^{14} \usk\gram\per\centi\meter\cubed$, in the right panel
$1.5\times 10^{10}$ and $8.1\times 10^{10} \usk\gram\per\centi\meter\cubed$. 
Vortex lines are cut off outside the outermost surface 
and beyond a length cutoff in order to limit cluttering.
The color indicates the vorticity magnitude, where lighter color corresponds 
to larger values.
The camera distance is the same in both panels.
}\label{fig:disk_vorticity}
\end{figure*}

For visualization purposes, we compute integral curves of the instantaneous vorticity.
We do not average the velocity field in time, because here the focus is on the impact of 
disturbances, not on the overall average fluid flow.
The result is shown in \Fref{fig:disk_vorticity}. We find that the vorticity
field is quite irregular outside the NS. Using an interactive rendering of the figure, we 
observed many closed vorticity lines, similar to those found in smoke rings. Our cursory 
visual inspection did not reveal any linked loops. Overall, the non-radial components 
dominate the vorticity. 

Comparing the disk before and after merger, we observe that the density
quickly becomes more axisymmetric after the NS collapses, as it ceases to inject spiral waves
into the disk. This can also be seen in \Fref{fig:remnant_xyt}.
The vorticity structure on the other hand does not become regular, i.e., 
the numerous small-scale vortices continue to dominate the local shear until the end 
of the simulation.

Our findings suggest a possible interpretation as follows. The rotating non-axisymmetric 
deformation and the radial oscillations inject a complicated 
pattern of waves into the disk that stir up the matter. 
The resulting  
perturbations dominate the vorticity on medium and small length scales.
As long as vorticity is conserved (which is not necessarily the case 
in hot matter and also not expected to hold exactly when using the curl 
as vorticity measure) one can expect the disturbances to manifest as closed 
vorticity lines.
Since vorticity lines are dragged along the fluid, the differential
rotation of the disk stretches small vorticity loops, resulting in
a predominantly non-radial orientation.   
Another possible interpretation would be turbulence. 

Whether the fluid is turbulent in the strict fluid dynamics sense or just
exhibits a very irregular looking flow, our results suggest that treating the 
flow inside the disk as laminar might not be sufficient
for all applications. Most notably, properties of the MRI instability are often
predicted in terms of rotation rate around the origin, density, and magnetic 
field strength. However, since the local shear is dominated by essentially 
random and time-dependent perturbations, this might not be justified. On the
other hand, a magnetic field of sufficient strength might have a dampening
impact on local vortices.

\subsection{Angular Momentum and Energy}
\label{sec:balance}

We now discuss the distribution of mass and angular momentum using the 
various measures discussed in \Sref{sec:diag}. Our main interest 
is whether the remnant collapses because of angular momentum transport 
within the fluid or because of angular momentum loss via GW radiation. We 
will answer this for the numerical results, but emphasize that angular momentum 
transport is most likely not captured correctly. The lack of effective magnetic 
viscosity might lead to underestimation of the dissipation of differential 
rotation, while the unavoidable numerical viscosity might lead to overestimation
in case of low actual viscosity. 

\begin{figure}
\includegraphics[width=0.95\columnwidth]{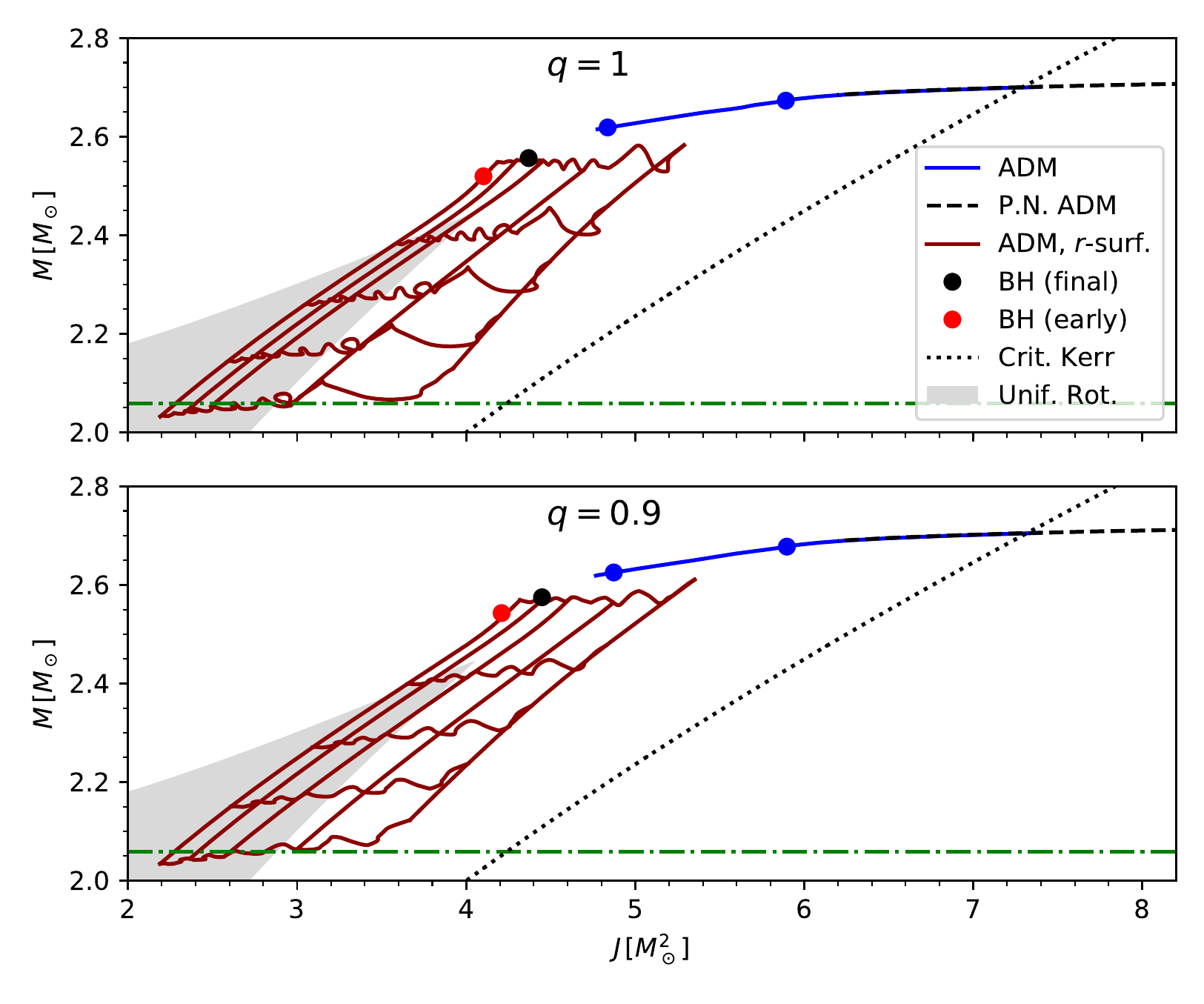} 
\caption{Evolution of energy and angular momentum for the equal mass 
system (top panel) and the $q=0.9$ case (bottom panel).
The solid blue curve shows total ADM energy versus angular momentum, and
the dashed black line a post-Newtonian approximation for the inspiral
of point particles.
The blue circles represent energy and angular momentum excluding the 
amount attributed to GW radiation outside $r=100\usk\kilo\meter$, at
times $1\usk\milli\second$ after merger and $1\usk\milli\second$ before
collapse.
The black and red dots mark energy and angular momentum of the BH
at the end of the simulation and $1\usk\milli\second$ after apparent horizon 
formation.
The diagonally oriented red curves show the contributions within 
spheres of constant radius to the ADM volume integrals, at regular 
intervals from $1\usk\milli\second$ after merger to $1\usk\milli\second$
before apparent horizon formation. The horizontally oriented red curves
show the time evolution of the contributions within spheres containing 
different fixed amounts of baryonic mass, ranging from $2.5\usk M_\odot$
up to the baryonic mass swallowed by the BH within 
$1\usk\milli\second$ after formation.
The shaded region is bounded by the mass shedding limit and smallest 
possible angular momentum of stable uniformly rotating NSs. 
The green
horizontal line marks the maximum mass of nonrotating NS.
The dotted line shows the angular momentum of critical Kerr BHs.
}\label{fig:evol_j_m}
\end{figure}

First, we establish how much angular momentum and energy is lost via GW.
The total ADM energy is shown in \Fref{fig:evol_j_m} as function of total
angular momentum. We mark the values at merger and collapse to visualize
the loss during the post-merger phase (the BH ringdown is negligible).
However, it would be wrong to directly associate this loss to the changes
in the remnant, for the following reason. At the time when the merger signal 
reaches the extraction radius, the space between remnant and extraction radius
contains strong GW radiation from the early post-merger phase (compare 
\Fref{fig:eqtov_evol}). We find that
the radiated energy and angular momentum corresponding to this part of the 
signal constitutes a significant fraction of the total loss. To get a better
handle on the energetics of the remnant, we computed the total energy and angular 
momentum at times when the wavefront passing a smaller sphere at time of merger
reaches the extraction radius. In the same way, we treat the time of collapse.
This is similar to using a small extraction radius,
but avoids extracting GW in the strong field zone.  
The resulting values for energy and angular momentum, which are also shown in  
\Fref{fig:evol_j_m}, should be more closely related to the changes within the 
remnant. We therefore think of those as energy and angular 
momentum of the remnant and disk.

For comparison, \Fref{fig:evol_j_m} also shows the ranges possible for uniformly
rotating NS with the initial data EOS, as well as the curve for critical Kerr BH.
Clearly, both the total energy and the total angular momentum of remnant and
disk (see above) are always larger than the maximum values for uniformly rotating 
NS. For the energy, this can be expected since the total baryonic mass
is in the hypermassive range. The comparison to the Kerr curve shows that
energy and angular momentum of remnant and disk could be realized by a BH 
at any time. 

The actual values for the final BH are shown in \Fref{fig:evol_j_m} as well.
The BH energy and angular momentum shown are computed using the isolated horizon 
framework and would correspond to the ADM values for an isolated BH. However,
the final BH is still surrounded by a massive disk, which accounts for the 
difference to the total ADM energy and angular momentum in the computational
domain. The figure also contains the values shortly after BH formation. 
The differences to the final values are mainly due to matter not in stable orbits
falling in during the first few $\milli\second$. The early BH is interesting because
it corresponds more closely to the part that actually collapsed.

In order to get an estimate of the angular momentum transport inside the remnant,
we study the integrands of the ADM volume integrals. At each time, we compute the 
contributions to ADM energy and angular momentum integrals within coordinate spheres
as function of the sphere's coordinate radius. \Fref{fig:evol_j_m} shows the resulting
curves at 5 different times. In addition, we show the time evolution for coordinate 
spheres with time-dependent radius chosen such that the baryonic mass within the 
radius stays constant. 
The time evolution of angular momentum and energy within those spheres is thus proportional to
the average angular momentum and energy per baryonic mass. 
The largest radius plotted is the one of the sphere that contains exactly the amount of baryonic 
mass that is swallowed by the BH within $1\milli\second$ of apparent horizon formation (but 
not exactly the same matter, as the swallowed region is non-spherical).

From the shape of the resulting grid, we deduce that the angular momentum loss
dominates the angular momentum redistribution within the remnant. Furthermore,
the angular momentum loss of the remnant is comparable to the loss by GW.
\Fref{fig:evol_j_m} in conjunction with our finding that matter is migrating 
outwards into the disk, as shown in \Fref{fig:radial_evol},  suggests that the 
associated angular momentum transport is subdominant to the GW in this case.

We recall that the integrands in the ADM volume integrals are gauge dependent 
quantities that depend on the time slicing. The use of coordinate spheres 
we used above also introduce a dependence on the spatial coordinates used in 
the simulation. It is therefore advisable to compare with other measures.

\begin{figure}
\includegraphics[width=0.95\columnwidth]{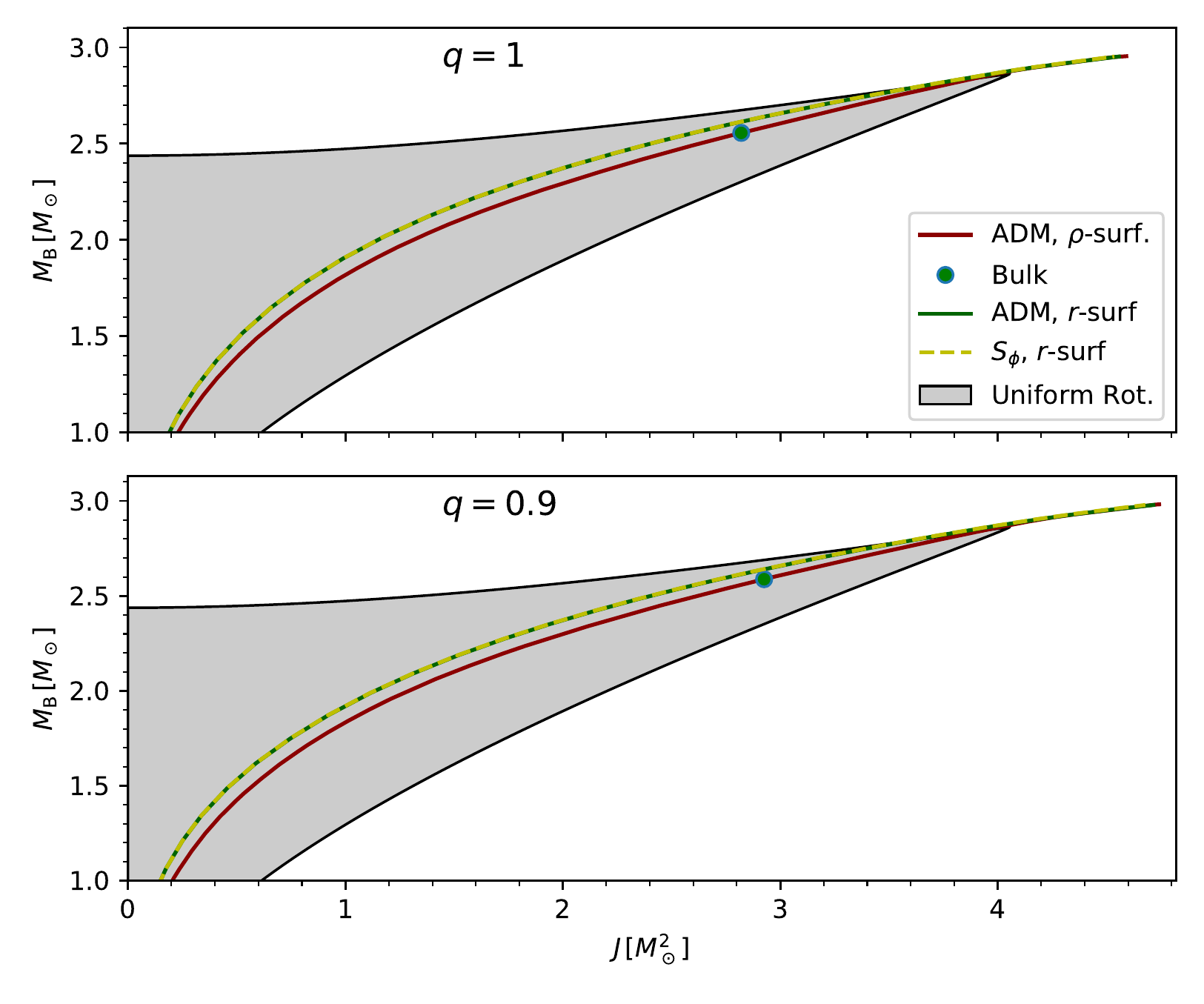} 
\caption{Distribution of angular momentum at time $1\usk\milli\second$ before
collapse for the equal mass system (top panel) and the $q=0.9$ case (bottom panel).
The solid red curve shows the contribution to the ADM angular momentum within
isodensity surfaces as function of the baryonic mass within the same surfaces.
The green dot marks the position of the remnant bulk (see main text). 
The solid green line shows the same for spherical surfaces (spherical with respect 
to simulation coordinates). It is hidden behind the dashed yellow curve, which shows
an estimate of the Komar angular momentum (see text). The shaded area shows the 
values possible for uniformly rotating NS.
}\label{fig:jadm_mass}
\end{figure}

One comparison we can do is between the ADM angular momentum and an approximation 
of Komar angular momentum (see \Sref{sec:diag}). The latter is dependent 
on the spatial gauge as well but in a different manner. The comparison
might reveal gauge dependencies of the results, but an agreement is no conclusive
proof that gauge effects are negligible.
The two measures are shown 
in \Fref{fig:jadm_mass} for the time shortly before collapse. 
We find that the Komar-type angular momentum measure matches the ADM
angular momentum almost exactly, as would be expected for an axisymmetric 
spacetime. 

Another simple comparison is between the ADM quantities computed within coordinate spheres
and those computed within the isosurfaces of mass density. The latter surfaces are independent
on the spatial gauge. Although the density distribution is not spherically symmetric,
we find that the two measures match well. Our comparisons indicate
that the qualitative picture we derived from \Fref{fig:evol_j_m} is not simply an
artifact of gauge effects.

For comparison, \Fref{fig:jadm_mass} also shows the allowed region for uniformly rotating
NS. Somewhat surprisingly, the remnant profile passes right through the uniformly
rotating model of maximum mass. We are unaware of a reason to expect such behavior, which might well be 
a numerical coincidence.

We are lead to the conclusion that the remnant studied in our simulations is 
driven to collapse mainly by angular momentum loss via GW, whereas angular 
momentum transport into the disk or within the remnant are less important 
factors.
We stress that our findings do not necessarily generalize to all HMNS. For
even longer lived remnants, the angular momentum transport can definitely 
become more important, as was shown in \cite{Nedora:2021:98}.

\section{Summary}

In this work we present a possible scenario for the fate of the merger remnant of GW170817.
We employ standard numerical simulation techniques and focus on the most fundamental
hydrodynamic processes, ignoring magnetic fields and neutrino radiation. The main motivation
is a better qualitative understanding of hypermassive merger remnants. To this end, 
we create novel postprocessing and visualization tools to analyze the numerical results.
Those methods provide a more detailed view on three key aspects of the simulated mergers.

First, we find that the merger remnants are not merely differentially rotating axisymmetric systems
deformed by some oscillation modes known from linear perturbation theory. 
The flow and the density deformation are best described in a rotating frame, where they
form a pattern that remains stable until the remnant collapses to a BH. 
A prominent feature of the flow pattern are secondary vortices in the outer 
layers. Such vortices were already observed in earlier studies restricted to the equatorial 
plane. Here, we visualized how they extend outside the equatorial plane.
The density deformation pattern in the aforementioned rotating frame is not a simple ellipsoidal 
deformation either. Instead, we find that the perturbation in the core is oriented nearly orthogonal 
with respect to the deformation near the transition zone to the disk. The latter
deformation seems related to the secondary vortices which are located at radii between the two
regimes. 

The deformation of the remnant is directly related to the GW signal. As in earlier studies,
we find that the GW frequency is twice the maximum rotation rate. The maximum rotation rate is 
modulated by a decaying quasi-radial oscillation, which matches exactly a modulation of the 
GW frequency. Another noteworthy aspect is that the spatial phase-shifts mentioned above imply 
cancellation effects in the quadrupole moment, and therefore the GW amplitude.
Moreover, the shape of the deformation undergoes a slow drift, such that cancellation effects 
become time-dependent.

One may speculate whether this effect can become pronounced enough to cause zero-crossings 
of the quadrupole moment in the rotating frame. Such crossings would explain secondary minima 
and phase jumps in addition to those occurring during merger, which are sometimes observed 
in numerical simulations of post-merger GW signals. For the examples studied
here, no secondary phase jumps were observed, however.

Second, we find at any time that the overall radial mass distribution in the remnant core 
is well approximated by profiles of a nonrotating isolated NSs. This fits well to the relatively 
slow rotation rate in the core, as was already observed in many earlier works. 
A key observation is that the profile directly before the onset of collapse matches the 
profile in the core of the maximum mass nonrotating NS.
In earlier work we proposed that, for generic hypermassive remants, collapse sets in 
exactly when the core reaches this critical density profile, which depends only on the EOS.  
The new examples add further support to this conjecture.

Third, we identify the mechanism responsible for the drift of the density profile leading 
towards the eventual collapse. By studying the radial distributions of mass and angular momentum 
and their time evolution, we rule out a change in differential rotation inside the 
remnant as dominant cause. We also rule out angular momentum transfer into the disk
and accretion onto the remnant. Instead, the main effect is the angular momentum carried
away by the strong GW emitted until collapse. 

We have to stress, however, that small
scale magnetic field amplification effects, which are not taken into account in our 
simulations, might lead to a large effective viscosity. This might speed up the loss of
differential rotation to a degree such that it becomes dominant over the impact of
angular momentum loss via GW.  Nevertheless, considering only the latter in a simulation 
provides an upper limit for the collapse delay that does not depend on effective viscosity.
We emphasize that the above statements refer to the systems studied here, which form
a hypermassive NS that emits a strong GW signal and collapses within tens of milliseconds.
Our results cannot be generalized to longer-lived remnants, which are not ruled out by
the observational data for GW170817. 

Another avenue for future research is the impact of neutrino radiation transport on the 
non-axisymmetry of the remnant. As in earlier work, we find a non-axisymmetric thermal
structure that corotates with the overall flow pattern. It is unknown how important  
the corresponding thermal pressure perturbations are for maintaining the non-axisymmetric
remnant perturbation. Since the latter are the cause of the post-merger GW signal, one
can speculate on a relation between neutrino cooling and the decay of the GW amplitude.

Last but not least, we investigate the disk surrounding the remnant and matter ejected 
from the system. The mass of the disk present in our simulation after BH formation is 
sufficient to allow the massive wind component (red component) inferred from kilonova AT2017gfo, 
although this would require an efficient mechanism for expelling matter. The dynamical mass 
ejection in our results is insufficient to explain the blue component inferred from the kilonova.

Making conclusive statements on the compatibility of our models with AT2017gfo, however, 
would require a convergence study with much higher resolutions. In fact, there is some 
tension with published results evolving our equal-mass model, including also neutrino 
radiation. This study does carry out a resolution test, but could not prove convergence 
of ejecta and disk masses for the model studied here either. Those simulations predict 
less dynamical ejecta and less massive disks. Noteworthy, they also result in shorter 
lifetimes of the HMNS. 

In our case, we find that matter is 
migrating into the disk from the HMNS, a result we also found in earlier work of different
systems. This effect might further increase when taking into account magnetically driven
winds from the HMNS remnant. We also find that the HMNS is strongly perturbing the disk,
which apparently causes part of the disk to become unbound. For the cases at hand,
the tidal ejection during merger is insignificant in comparison.

The above observations suggests that the lifetime of the HMNS remnant --- which is extremely 
sensitive to the total mass and to numerical errors because the system is close to collapse --- is indeed one of 
the main uncertainties regarding disk mass and mass ejection. It could prove difficult to 
find analytic fits to the parameters of the binary, and it might be advisable to treat the lifetime 
of HMNS as a free (albeit constrained) parameter in such fits.

The perturbation of the disk by the HMNS also has an effect on the velocity field in 
the disk, as shown by novel 3D visualizations of the vorticity field both 
before and after BH formation. We 
observe an irregular vorticity field instead of the ordered structure that would 
be present for a Keplerian velocity profile. This indicates that the shear on medium 
length scales is dominated by the disturbances originating from the remnant. 

The irregular vorticity structure is relevant with regard to estimates for 
the time and length scales of magneto-rotational instabilities, because the analytic 
models used for such predictions are based on disks with an orderly flow.
Although the density perturbations quickly settle down after BH formation, we find 
that the vorticity remains irregular until the end of the simulation. 

In this work, we focused on two examples only and refrained from costly high-resolution 
studies. Those two examples paint a qualitative picture of the HMNS structure and evolution.
As a future step, the analysis developed in this work needs to be applied to more simulations
in order to determine which parts of this picture are generic. This will also benefit
the development of more realistic models for HMNS created in mergers. Such models 
are needed for the analysis of future GW detections of a post-merger signal, as performing 
a large number of brute force merger simulations is computationally too expensive.
As qualitative examples for calibrating such models, we provide the GW data from our 
simulations.

\acknowledgments
This work was supported by the Max Planck Society's Independent Research Group 
Program.
The numerical simulations and renderings were performed on the 
\texttt{Holodeck} cluster at the Max Planck Institute for Gravitational 
Physics, Hanover. The authors thank Tim Dietrich for helpful comments on the 
manuscript.
\bibliographystyle{article}
\bibliography{article}

\end{document}